\documentclass[prl,showkeys,twocolumn]{revtex4-1}

\usepackage{color,graphicx,amssymb,amsmath}
\usepackage{hyperref}
\usepackage[pagewise]{lineno}

\begin{document}
 
\newcommand{\ch}[1]{\textcolor{green}{\textbf{#1}}}

\title{How ants move: individual and collective scaling properties }
 
\author{Riccardo Gallotti$^{1,2}$}
 \email{rgallotti@gmail.com}
\author{Dante R. Chialvo$^{2,3}$}
 \email{dchialvo@conicet.gov.ar}
 \affiliation{$^{1}$ Instituto de F\'isica Interdisciplinar y Sistemas Complejos (IFISC), CSIC-UIB, Campus UIB, ES-07122 Palma de Mallorca, Spain}
 \affiliation{$^2$ Center for Complex Systems $\&$ Brain Sciences (CEMSC$^3$), Universidad Nacional de San Mart\'in, 25 de Mayo 1169, San Mart\'in, (1650), Buenos Aires, Argentina.}
 \affiliation{$^3$ Consejo Nacional de Investigaciones Cient\'ificas y Tecnol\'ogicas (CONICET), Godoy Cruz 2290, Buenos Aires, Argentina}
 
\begin{abstract}

The motion of social insects is often used a paradigmatic example of complex adaptive dynamics arising from decentralized individual behavior. In this paper we revisit the topic of the ruling laws behind burst of activity in ants. The analysis, done over previously reported data, reconsider the  causation arrows, proposed at individual level, not finding any link between the duration of the ants' activity and its moving speed. Secondly, synthetic trajectories created from steps of different ants, demonstrate that a  Markov process can explain the previously reported speed shape profile. Finally we show that as more ants enter the nest, the faster they move, which implies a collective property. Overall these results provides a mechanistic explanation for the reported behavioral laws, and suggest a formal way to further study the collective properties in these scenarios. \\
Citation: Gallotti R, Chialvo DR. 2018 How ants move: individual and collective scaling properties. J. R. Soc. Interface 15: 20180223.  DOI: 10.1098/rsif.2018.0223
\end{abstract}


\maketitle
 
\section{Introduction}

The behavior of social insects constitutes beautiful examples of adaptive collective dynamics born out of apparent purposeless individual behavior \cite{Wilson1971,Holldobler1990}.  The richness of the phenomena offers plenty of opportunities to test theories dedicated to understand fascinating aspects of collective organization and behavior.  An emergent issue is to dissect the dynamics that are routed on individual versus those explained by collective forces. (A recent review can be found on  Ref. \cite{Feinerman2017}).  A relevant perspective allowing to point out the effects of collective behavior is the analysis of insects' speed. For instance, the walking speed of an ant changes after interaction with other ants \cite{Razin2013}, where the change depends on the speed of the second ant. Consequently, it has been suggested that a building up in speed, following the interaction with fast returning foragers, might be a the basis of an ant's decision to leave the nest for a foraging trip  \cite{Razin2013,Davidson2017}.

Recent work \cite{Christensen2015,Hunt2016} found that ants' bursts of activity moving both inside and outside their colony's nest exhibit a power-law relation between the duration of the activity and its average speed. Successive motion events, defined as the segment of data between two consecutive motionless instances, was found to obey a universal speed shape profile. The authors concluded  that this predictability implies that the duration of each ant's movement  is ``somehow determined before the movement itself'' \cite{Hunt2016}, thus placing important weight into the individual ant's  spontaneous behavior.

In this work we revisit the topic proceeding to analyze the same data sets as in \cite{Christensen2015} to disambiguate the individual properties that can be judged as cognitive from the features that can be potentially based on other aspects of the animal's physiology.
A null model is introduced allowing to create synthetic trajectories by adding steps of different ants selected at random. The results of the analysis suggest an alternative explantation simpler than the proposed causation arrows between the ants' duration of activity and its speed suggested in Ref.~\cite{Christensen2015}, demonstrate that the true origin of the universal shape profile of the speed can be reconnected to a Markov process taking into account the observed ants' speed auto-correlation that is based in the animal's physiology.

Finally we show a novel and distinctive collective effect by which as more ants enter the nest, the faster they move.

\section{Methods}

{\bf Data.} The data re-analyzed here was already described in \cite{Christensen2015}, and correspond to \emph{Temnothorax albipennis} ants, which forms small colonies in rock crevices, approximated in the laboratory by a 100$\times$100 mm$^2$ Petri dish where food and water are available at will.  Ant workers were identified by markers with unique combinations of color paint dots. Ants' individual trajectories were video recorded within two nest sizes: 35$\times$28 mm$^2$ and 55$\times$44 mm$^2$ in a randomized order from each of three colonies (labeled $C^1$, $C^2$ and $C^3$, and indexed according with the nest sizes used). Ants' trajectories were reconstructed manually from the video with a cursor with a time resolution of 0.1 secs. To compare with previous work \cite{Christensen2015,Hunt2016}, we use similar preprocessing steps. The data was coarse-grained to an average sample interval of 0.8  secs in order to reduce the very small-scale fluctuations. An ``activity event'' was defined using a speed threshold $v_e =0.001$ mm/sec, below which it said that the ant is inactive.

Typical ants' trajectories are plotted in Figure 1. Figure 2 shows examples of the speed changes for a group of ants during the first minute recorded. It can be seen that the ants' motion is discontinuous and irregular, eventually stopping and restarting their walk. Indeed, decreases in the ant's speed are related with the so-called ``marking behavior''  in which the ant touches the tip of its gaster to the surface, a known property in various ant species \cite{Beckers92,Czaczkes2013,Holldobler78,Holldobler81}.
\\
\\
{\bf Model.} The causality argument proposed in \cite{Christensen2015,Hunt2016} is supported by the comparison with what we can call a {\it zeroth order} null model, where synthetic realizations of the ants' trajectories were to be made by randomly extracting speeds from the distribution $P(v(t))$ thus ignoring the possible correlations with the current state of the movement. 
  
Here, we propose an improved {\it first-order} model taking into account of the fact that autocorrelation plays a strong role in animal movement processes. The model is based on the statistical reconstruction of return map for the ants' speeds. This map can be observed in Fig.~3A, where we plot for all time series, all pairs of consecutive values of speeds $v(t), v(t+1)$ (dots in Fig.~3A). Subsequently, values of $v(t)$ were binned and its corresponding $v(t+1)$ averaged (circles in Fig.~3A).  
Similarly, to construct the synthetic speed trajectories of the null model, we define a conditional probability distribution for the subsequent velocities $P(v(t+1)|v(t))$ from the data with bins of $\Delta v = 0.1$ mm/sec. A special bin is used for speeds after rest events (that is, speeds below the threshold $v_e = 0.001$ mm/sec). 
Then, we describe an ant's speed as a Markov process which draws from the binned distribution, using the following iterative procedure:
1) An initial value of speed $v(t)$ is randomly chosen from the bin of speeds following a rest event. 2) The next speed is determined by randomly choosing a $v(t+1)$ inside the corresponding bin of $P(v(t+1)|v(t))$ 3) Such value is then considered as the new $v(t)$ for the new iteration. 4) If $v(t+1)<v_e$, the movement is considered completed. 5) When a movement is finished, a rest time $\tau$ is extracted from at random from the experimental data. 6) After a time $\tau$ with no movements, we restart from the first step.

\begin{figure}
 \centering{\includegraphics[width=.45\textwidth]{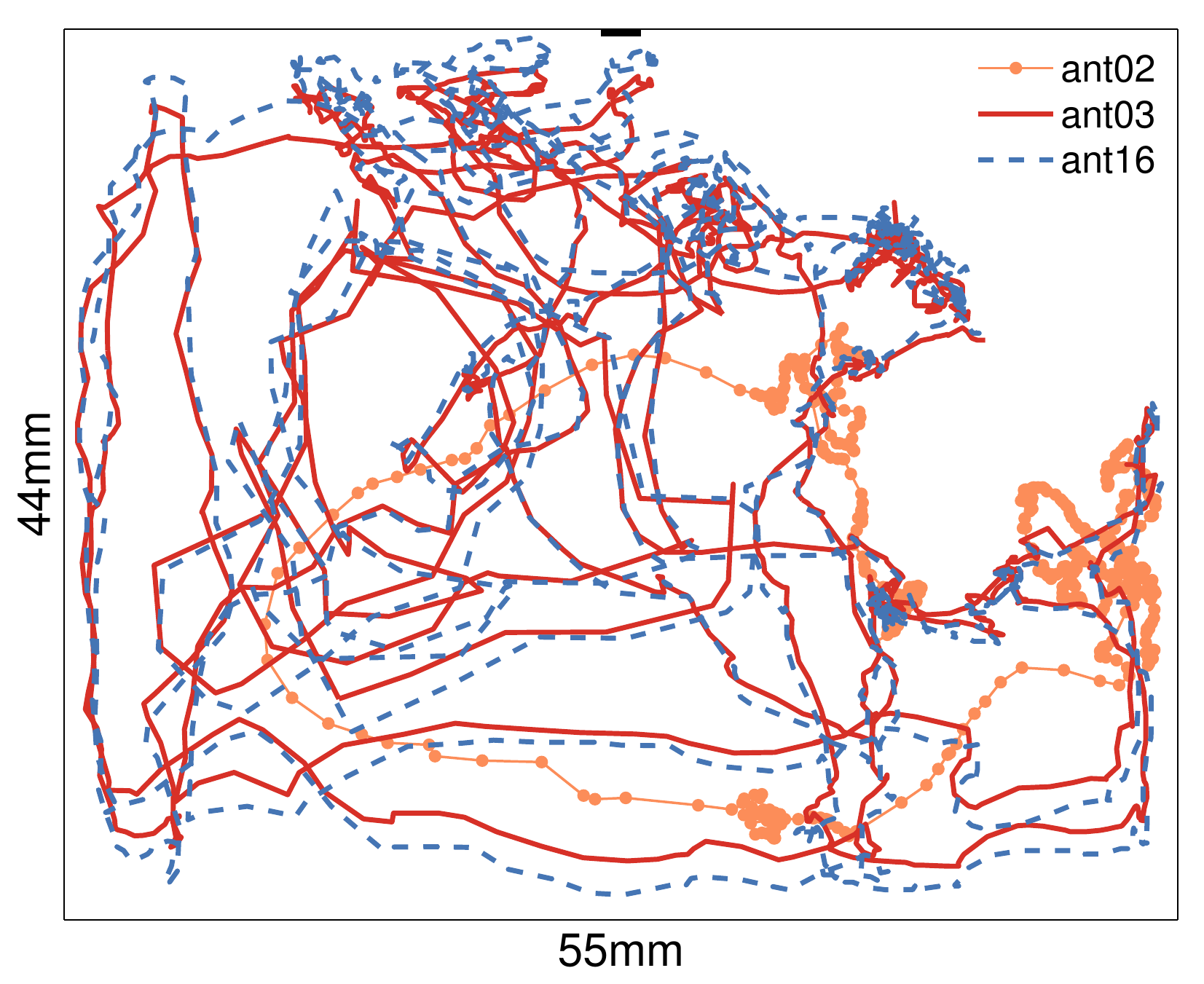}}
\caption{Examples of typical ants' trajectories:  three ants from colony $C_{55\times 44}^2$, after entering the nest from the point  marked with a small bar (middle region of the top side). Notice the movements of two of them (ants 03 and 16) which are manifestly correlated.}
\end{figure}
\section{Results}

The results are organized as follows: first we describe the observation that can be derived by analysing the map of discrete consecutive speeds from the raw data. This map is used to define the {\it first-order} null model, which is relevant to isolate the role of auto-correlation in the individual ant's dynamic. After that, we compare the scaling statistics for both processes: the recorded and null model simulated ants' trajectories, including the events' duration, speed, intervals and shape. Finally we describe a novel dependence of speed as a function of the number of ants moving.

Despite the wide scatter of the data points in Figure 3A, the {\it average} $f (v_{(t)})$ (circles and continuous line) make evident the auto correlation between consecutive steps.  Thus, the average dynamics of the ants' speed can be, in part, be described by the average map $f$; i.e., values larger than $\stackrel{*}{v}$ (arrow) tend, on the average, to decrease (i.e., the average value stays below the identity line). For values of speed smaller than  $\stackrel{*}{v}$ there are two tendencies: if the ant is moving then it will speed up in the next time step.  For the case that the ant is not moving,  we cannot say anything about {\it when} it will start moving (although see below); nevertheless, whenever it does it, it will do it with an average speed of the order of $\stackrel{*}{v}$ as indicated by the circled symbol at the origin of Fig. 3A.
 
\begin{figure}[ht]
\centering{\includegraphics[width=.5\textwidth]{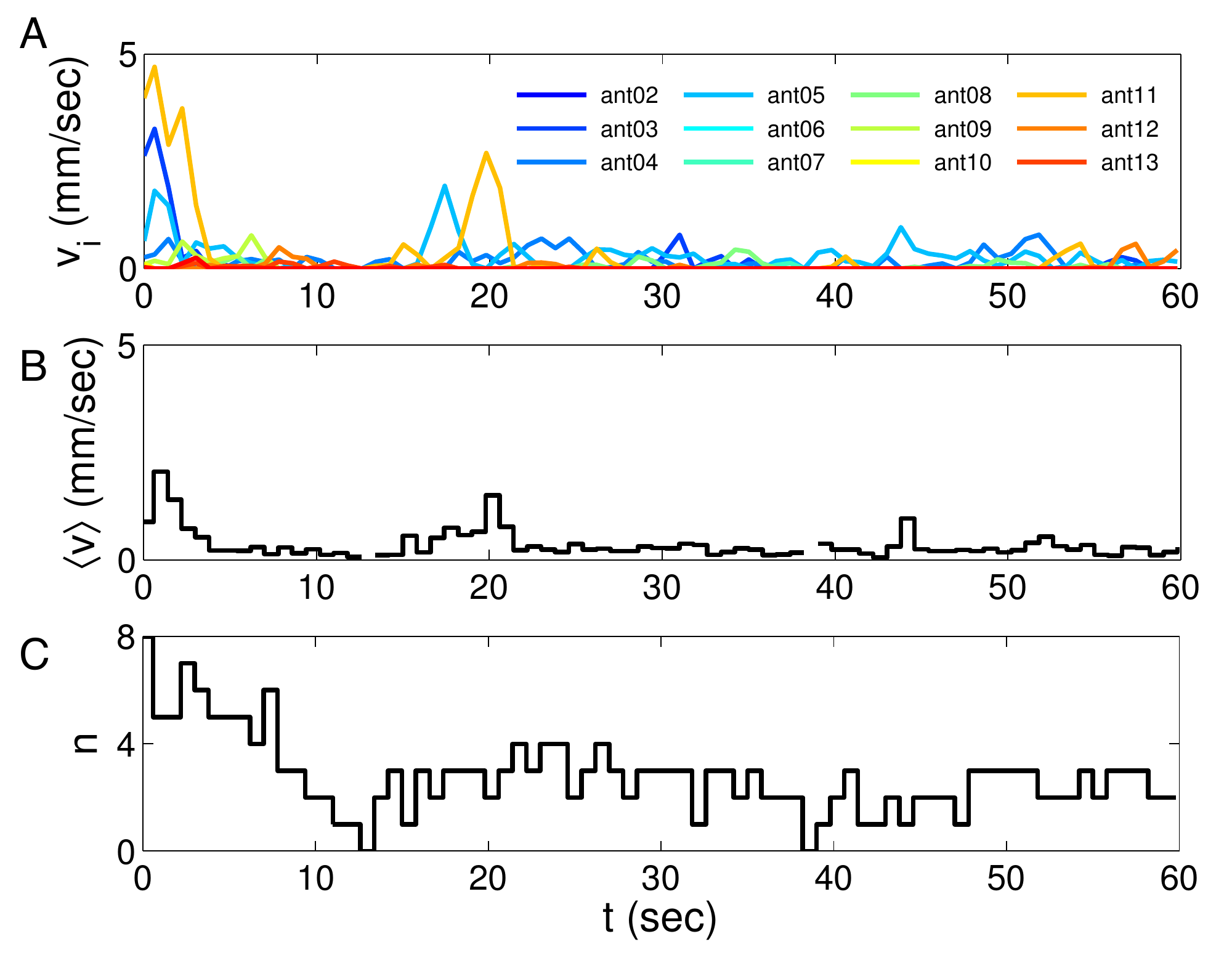}}
\caption{(a) Speeds of the ants moving in colony $C_{55\times 44}^1$ during the first recorded minute.  (b) Whenever at least a single ant is moving we can define the average speed $\overline v(t)$. (c) The number $n$ of moving ants. }
\end{figure}

The map $f$ in Fig. 3A can be used to test the potential origin of the fluctuations in the ants' speed,  constructing time series of speeds following a given null hypothesis. Following \cite{Christensen2015,Hunt2016} the variables of interest are the individual bouts of motion, defined by the speed fluctuations of a given ant, starting and ending with immobility (see Methods). For instance, Figure 3B shows a real event (circles and continues line) lasting 20 secs. The other six traces (dashed lines), starting at 4 secs, are stochastic realizations constructed using the map of panel A. Specifically, time series of speeds were computed iteratively: at each step the next speed value $v_{(t+ 1)}$ is calculated  as follows: given a speed $v_{(t)}$, the next speed is  randomly selected  from  the  subset of $v_{(t+1)}$ values corresponding to the bin $v_{(t)}- \Delta v < v_{(t)} < v_{(t)} + \Delta v$ with $\Delta v=0.05 ~$mm/sec). The computed value became the new current speed and the iteration proceeds.  
It can be seen, in the example, that the trajectories of speeds constructed this way varies both in size and duration. It is important to note that the null model constructed in this way breaks some, but not all, correlations present in the real ants' walks, since the map in Figure 3A preserves the average serial correlations in the speed.

As will be discussed later, the stochastic procedure used in our {\it first-order model} shuffles, in a sense, the ``decisions'' of individual ants at each time step. This deletes any predetermined $\it individual$  plan  (as argued in \cite{Hunt2016}) about how long and fast it would be any given bout of motion, while maintaining some of the properties of the speeds auto-correlation (see Supp. Info.).
Very different results are observed (see Supp. Info.) if the synthetic realizations were to be made with the {\it zeroth-order} null model by randomly choosing speeds regardless of their preceding one, as done in Ref. \cite{Christensen2015}.

\begin{figure}[htbp]
\centering{\includegraphics[angle=0,width=0.35\textwidth]{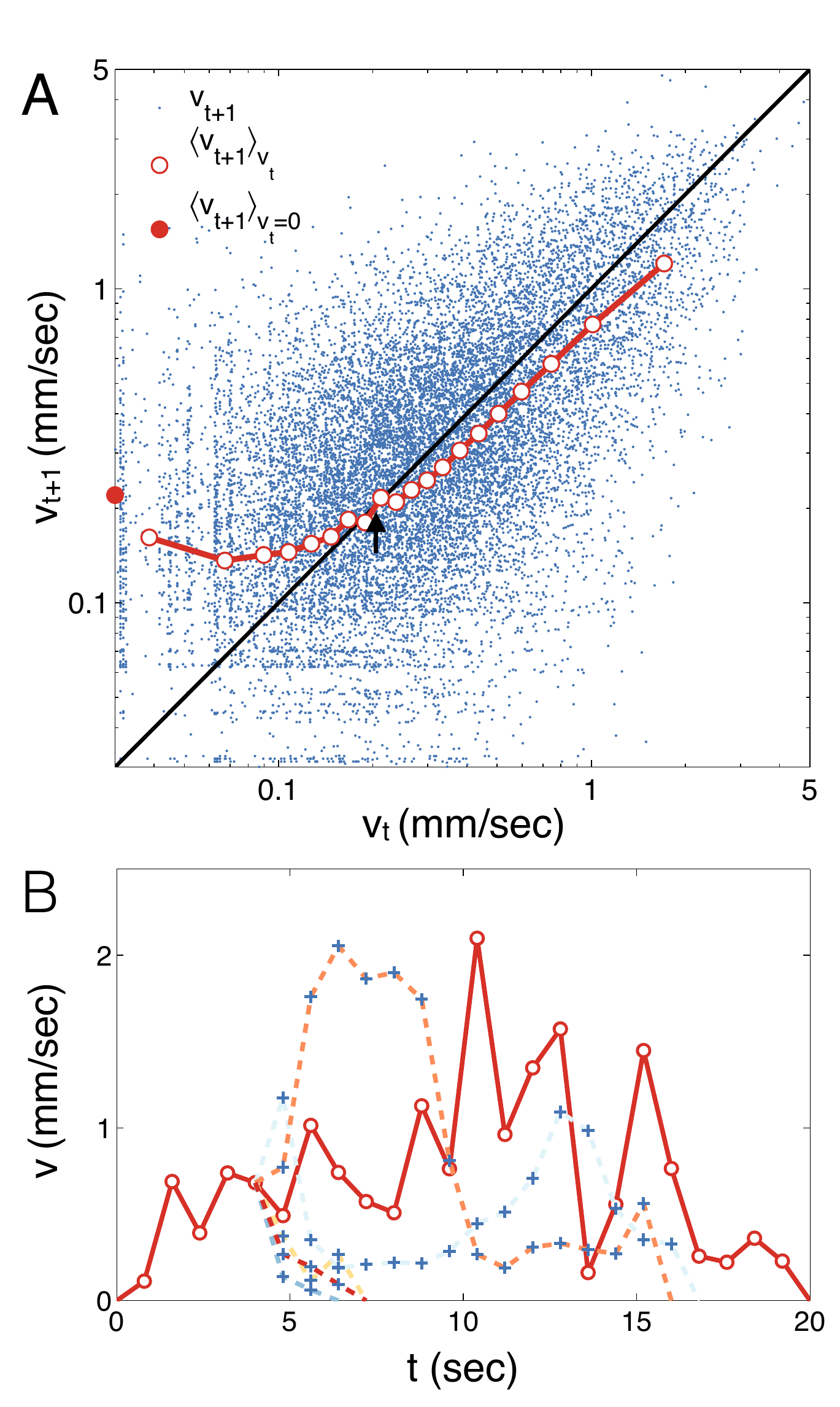}}
\caption{Extracting the null model  map. (A) Return map extracted from the raw data by plotting consecutive $v(t)$ samples (dots) and its binned average $f (v_{(t)})$  (circles and continuos line) over-imposed (please note the log axis). (B) An example of a real event  $v(t)$ (red solid line with circles) and six synthetic trajectories (dashed lines). In this example, differently from the implementation in our simulations, the trajectories are the same until $t=4$ sec. Then they evolve independently following our model starting from the real trajectory with initial condition $v_0 = v(4 \mathrm{\ sec})$ and selecting subsequents steps from the map in panel A. Notice that synthetic trajectories vary both in size and duration. Data from colony $C_{35\times 28}^2$, the scaling for other colonies are displayed in the SI.}
\end{figure}


Armed with an appropriate null model, now we can proceed to analyze the ants' data and to compare its statistics with synthetics time series. Of interest here are the ``events'', which are defined as the bouts of activity starting and finishing on a quiet instance (recall that quiet was defined as any speed below the $v_e$ threshold as in \cite{Hunt2016}).

First we compute the densities of event's duration T, the event's size S (that is, the total distance covered during the motion event), and their mutual relation, which are plotted in Panels A-C in Figure 4 respectively.  
Overall, the results shows that  the null model agrees very well to replicate  more than the 95\% of the real events, while fails to replicate the empirical probabilities of the longest lasting events (of the order of $1e^{-4}$ $<$ P $<$ $1e^{-2}$).  This disagreement might be a consequence of the higher order correlations which are not taken into consideration in our {\it first-order} model. Indeed, as we see in the Supp. Info, the empirical auto-correlation function decays slower than in our model, which can possibly lead towards longer moves.  

\begin{figure}[htbp]
\centering{\includegraphics[width=.4\textwidth]{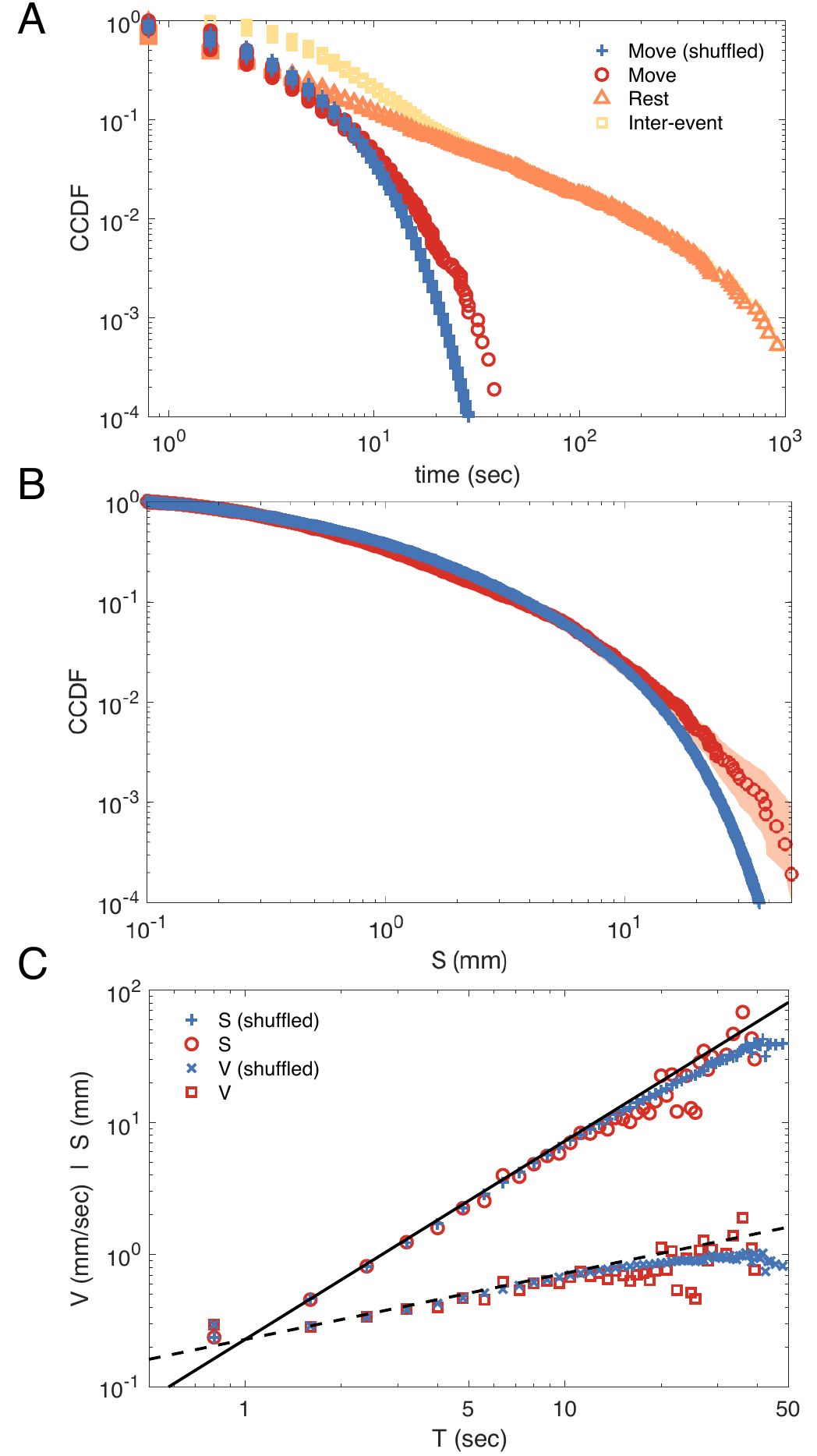}}
\caption{Real and null model trajectories exhibit very similar scaling statistics. (A) The complementary cumulative distribution function of the events duration $T$ , the inter-event time and rest time, for shuffled trajectories and real ants. (B) The complementary cumulative distribution function of the events size $S$ for real ants and the null model. The orange shaded area indicates the 95\% confidence interval for the synthetic CCDF computed via bootstrap. The difference between the two curves is restricted to the very few relatively long events. (C) The scaling of the events size and speed as a function of  events duration for the raw data as well as  for the null model shows no relevant differences. Continuous and dashed lines correspond to slopes of 3/2 and 1/2 respectively. Data from colony $C_{35\times 28}^2$, the scaling for other colonies are displayed in the SI, where we also report a dependence in the numerical value with the nest size already pointed out in~\cite{Christensen2015}. In the simulated data we generated 1 million synthetic events.}
\end{figure}

 Looking at the scaling relations in Figure 4C, is apparent that the exponents could be consistent with a stochastic process  where changes in speed are given by an uncorrelated noise
$\Delta v(t+1) =v(t+1) - v(t) = \xi(t)$, which would yield $v(T)\propto T^{0.5}$ and therefore $S(T)\propto T^{1.5}$ \cite{baldassarri:2003}. However, the fact is that {\it consecutive} speed samples of the raw data are correlated, as demonstrated by the average map of Fig. 3A. Because of that, it is of less interest to consider 
the zeroth-order null model in which these correlations are ignored and the time series are built by random shuffling of each raw time series of speeds, where speeds become independent by duration (see SI and Fig. 8 of Ref. ~\cite{Christensen2015}) 
\begin{figure}[h]
\centering{\includegraphics[angle=0,width=0.367\textwidth]{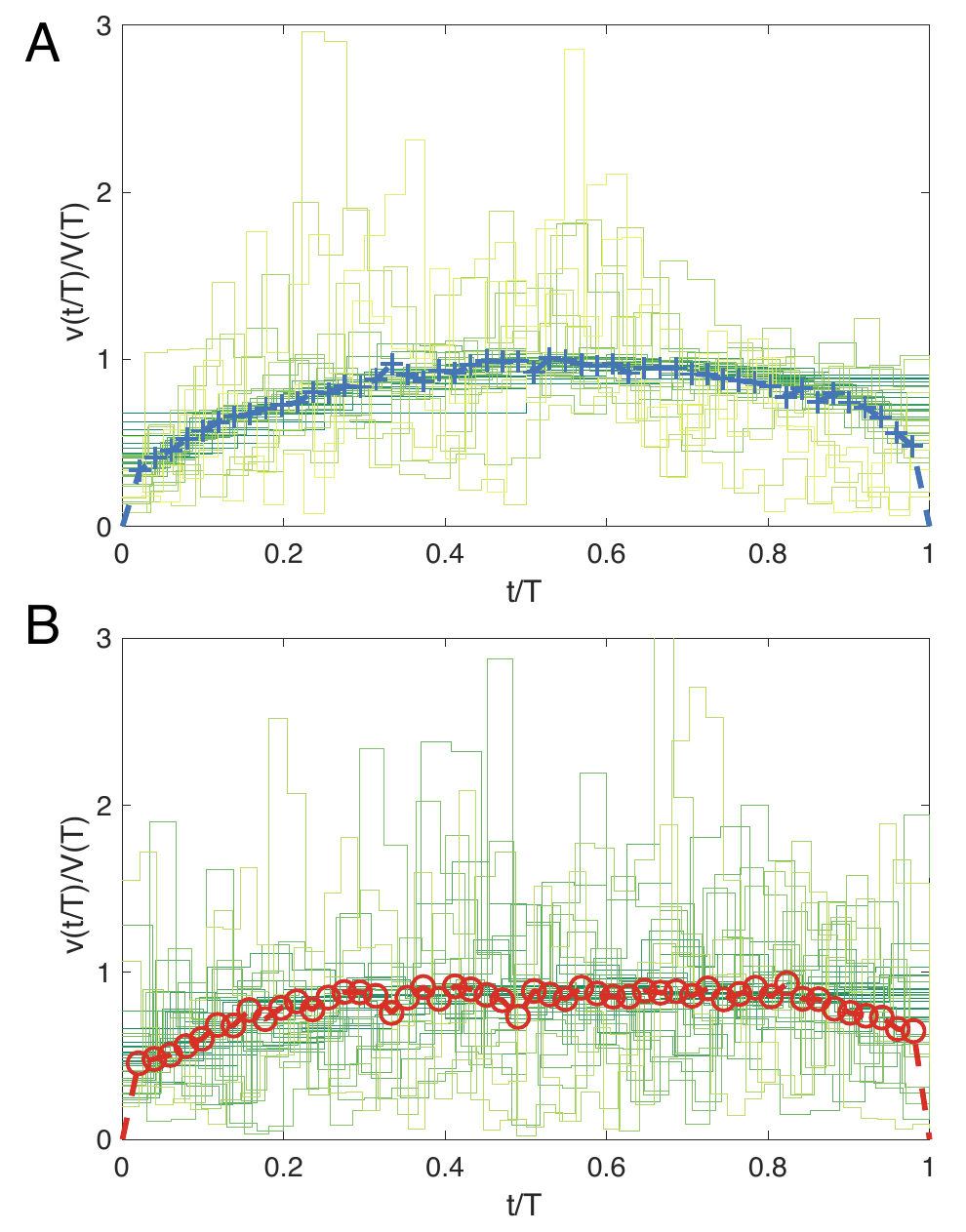}}
\caption{The normalized average speed profile for shuffled trajectories (panel A) and real ants (panel B) for colony $C_{35\times 28}^2$, other colonies are displayed in the SI. As observed previously in Ref~\cite{Christensen2015}, after rescaling, all the real events (panel B) profiles collapse to an universal function that presents a plateau close to the average speed at the middle of the trajectory. The same collapse can be observed in the synthetic trajectories in panel A, that also display a similar skew (see SI) even if the shape of the curve is less flat.}
\end{figure}

\emph{Shape collapse:} The functional  relation between the lifetime of an event and its size
shown in Figure 4C, observed in both, real and null model, implies  that the time series of speed is self-affine \cite{Mandelbrot1985}. Thus, after appropriate rescaling,  an average shape descriptive of the events shall be extracted, as was reported earlier for this data \cite{Christensen2015}, as well as for human activity \cite{Chialvo2015}.
We observe in Figure 5 that equally  satisfactory collapse of the trajectories can be obtained in both cases, the null model (top panel) and the raw data (bottom panel). However there is a small, but relevant, difference. While the shape function of the model  resembles more to the expected inverted parabola, the one derived from real walks exhibits a plateau.  A less noticeable feature, but common to both cases is the presence of a slight asymmetry. These similarities and differences will be discussed later on.
\begin{figure}[h]
\centering{\includegraphics[width=.4\textwidth]{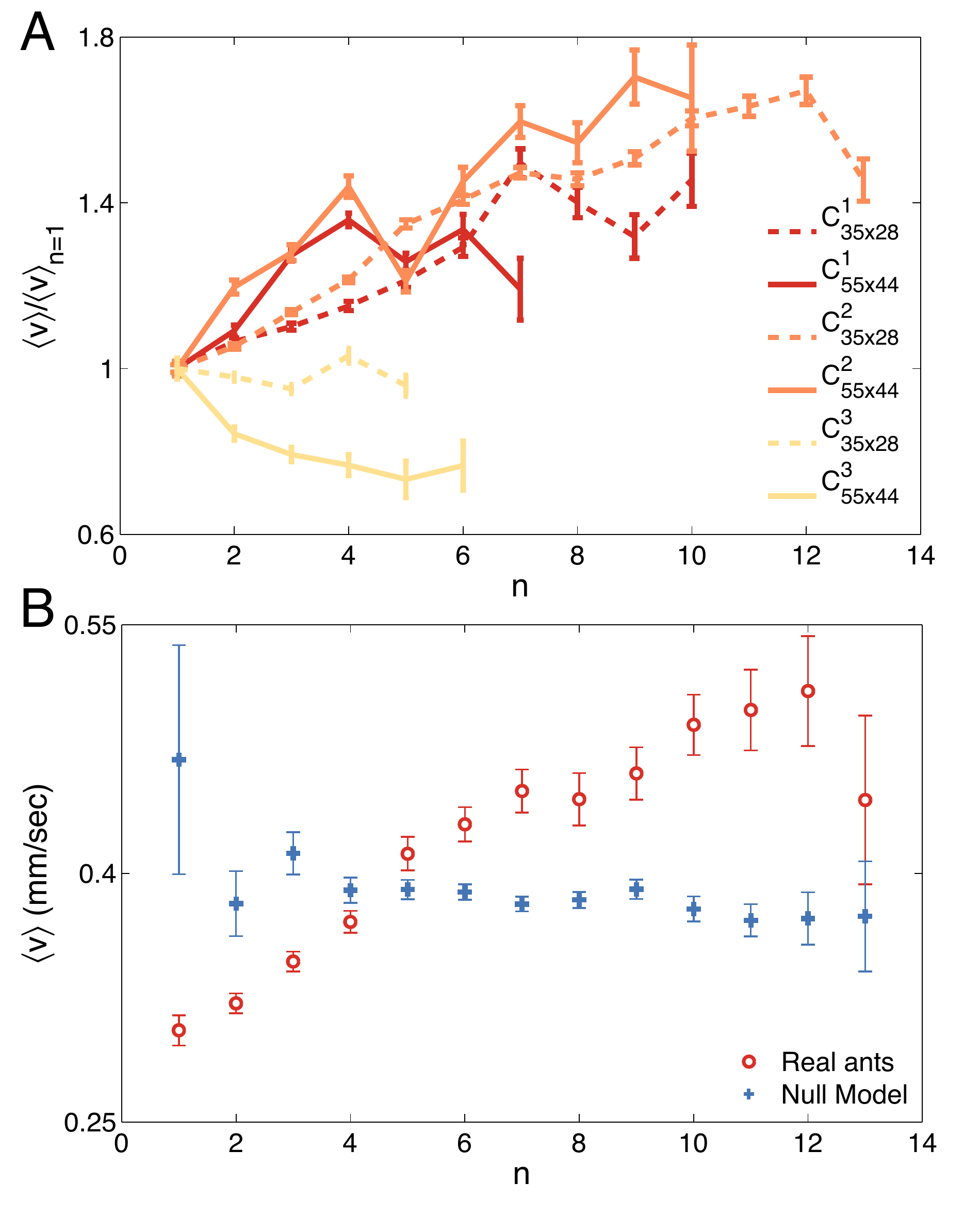}}
\caption{``The more the faster". (A) When more ants are in movement, they move faster. We observe that for the two larger colonies, $C^1$ and $C^2$, $\langle\overline v\rangle_n$ grows with $n$.  (B) The null model shows that growth of speed with $N$ is not due to single ants' behavior (Data from colony $C_{35\times 28}^2$, other colonies are displayed in the SI).}
\end{figure}

The results so far indicate that the scaling relation and the characteristic shape or the bouts of ant's motion are indistinguishable from the results of the null model trajectories in which {\emph consecutive speed increments} are shuffled at \emph{each} step among different ants and moments in time. This similarity is one the main results and clearly goes against the interpretation given in \cite{Hunt2016} when stating that ``ants determine their next move at rest'' assigning a cognitive root to the observed universality. To remark the contradiction, notice that the null model can be visualized ``as if'' at each step each ant uses another randomly chosen ant's decision for its own next step. The observed universal characteristics of ants' movement can be reconstructed using as only ingredient the auto-correlations of speeds, which origin is probably to be found in the animal's physiology.

\emph{Speed dependency with N:}  We observed, in the two colonies with the largest number of ants (but independently from the nest size, see Supp. Info.), an additional effect concerning the dependency of speed as a function of the number of ants moving.  Panel A of Figure 6 shows that as this number increases, the average speed also increases. Specifically, at each time $t$ we count the number $n$ of ants in motion,  and we compute their average speed $\overline v(t)$. The quantity $\langle\overline v\rangle_n$ represents the average speed of moving ants in a moment where $n$ ants are moving. 
To disambiguate  the origin of this dependency, we compared the growth in speed observed for the 22 ants tracked 
in colony $C_{35\times 28}^2$  with the same analysis performed on the same number of artificial time-series obtained from our null model. It is seen (Panel B of Figure 6)  that the speed for the  artificial trajectories constructed with the null model is  independent from the number of ants moving. In this way we can safely exclude the possibility that it is an artifact related to the larger average speed of longer trajectories, instead the results argue in favor of a effect emergent from the ants' interaction. 

\section{Discussion}

The main findings can be summarized along three points. First, from an individual perspective, the similarity between the {\it first-order} null model and real ants results takes considerable weight from the previous suggestion indicating that the duration of each ant's movement is ``somehow determined before the movement itself'' \cite{Hunt2016}. 
The proposed causation arrows in previous reported  is not found in our results, as the universal shape of the $v(t)$ curve is not a behavioral feature but the necessary consequence in movements characterized by random accelerations integrated over time.
 
Indeed, at {\it each} step of the synthetic data, ant $i$ future speed is determined by the decision made in another moment by another ant $j$, which completely ignores ant $i$'s past history or determination. It is clear that under such conditions any informed plan or determination taken by ants is hard to sustain.

Second, despite the similarities, it is shown that the average shapes of the individual movement events exhibits a small but relevant difference respect to the synthetic data. Figure 5B shows that the  shape of the empirical events is not completely parabolic as in Figure 5A , which is expected from a random walk. It is, instead,  rather flat.
This type of plateau, according with the analytical results of Baldassarri \cite{baldassarri:2003}, is expected for a particle executing a random walk on a potential well, in which the restoring ``force'', depends on the value of $v$. In simple words, relatively large increments of $v$ are harder for larger absolute values of $v$, something that make also sense from a biological standpoint.  In passing notice  that the slight asymmetry  is not expected for the fully stochastic memoryless linear process of Ref. \cite{baldassarri:2003}.

Third, the analysis unveiled a new observation in this type of data, dubbed here ``the more the faster'' effect; as 
more ants are in motion in the arena, the faster they move. This result is clearly a collective property, which is consistent with a number of theoretical ideas on trail formation, where at a certain density of agents directed motion (and subsequent individual' speed increase) emerges \cite{Rauch1995,Chialvo1995}.  It is our hope that this observation would encourage further experimental studies on the link between speed and colony dimensions, with the particular caution of forming differently sized groups of ants from the same colony, instead of from different colonies, to control for variation among field colonies.

Overall, these results provides a mechanistic explanation for the reported behavioral laws, and suggest a formal way to further study the individual and collective properties of animals trajectories. In particular, the modeling approach we proposed, based on the description of movement patterns as a random process characterizing the evolution of speeds in time, can be enlightening  for the open discussion on L\'evy Flights as models of animal movement~\cite{Edwards2007,Pyke2015,Reynolds2015}.

The L\'evy Flight hypothesis, proposing a universal scale-invariant character of animal motion, does not account for the interaction between different animals \cite{Breed2015} and for the autocorrelation in consecutive moves~\cite{Augermethe2015}. A deeper quantitative understanding of animal movement behavior can be gained by going beyond the simple characterization of movement patterns in terms of step-length distributions~\cite{Reynolds2018}. A model of randomly accelerated walkers, describing the evolution of speeds as acceleration kicks at random times, already proved successful in showing that the long-standing interpretation of human displacement as L\'evy flights was incorrect~\cite{Gallotti2015}.

The spontaneous movement of animals should necessarily be seen as the product of a continuous decision making process, adjusting the organism's behavior in response to both past and present events~\cite{Pyke2015}. This perspective requires accurate analysis of the experimental data, which must be recorded consistently with the spatial and temporal scales of the organisms' movement decisions \cite{Plank2009,Gallotti2018}.

\section{Data accessibility}
\noindent All data is accessible from the authors upon request. Ants' data have been already shared in \cite{Christensen2015} and is available at https://data.bris.ac.uk/
data/dataset/cmcs6znssfim12zo6zzmur1hq
\section{Competing interest}
\noindent Authors declare no competing interests.
\section{Authors' contributions}
\noindent DRC and RG conceived the study,  RG and DRC participated in data analysis, DRC and RG wrote the manuscript.  All authors gave final approval for publication.

\section{Acknowledgements}
\noindent RG acknowledges funding from the SESAR Joint Undertaking under grant agreement No. 699260 included in the European Union's Horizon 2020 Research and Innovation Programme. DRC thanks support from CONICET and Universidad Nacional de  San Mart\'in, Argentina.

\clearpage
\pagebreak

\onecolumngrid
\appendix

\section*{\large Supplementary Information}
\section*{Autocorrelation}
While the zeroth-order model is totally blind to the speeds autocorrelation, our first-order model ignores higher level correlations.  To quantify correlations in time, we study here the autocorrelation of speed for ants in colony $C_{35\times 28}^2$.

\begin{figure*}[h!]
\centering{\includegraphics[width=.6\textwidth]{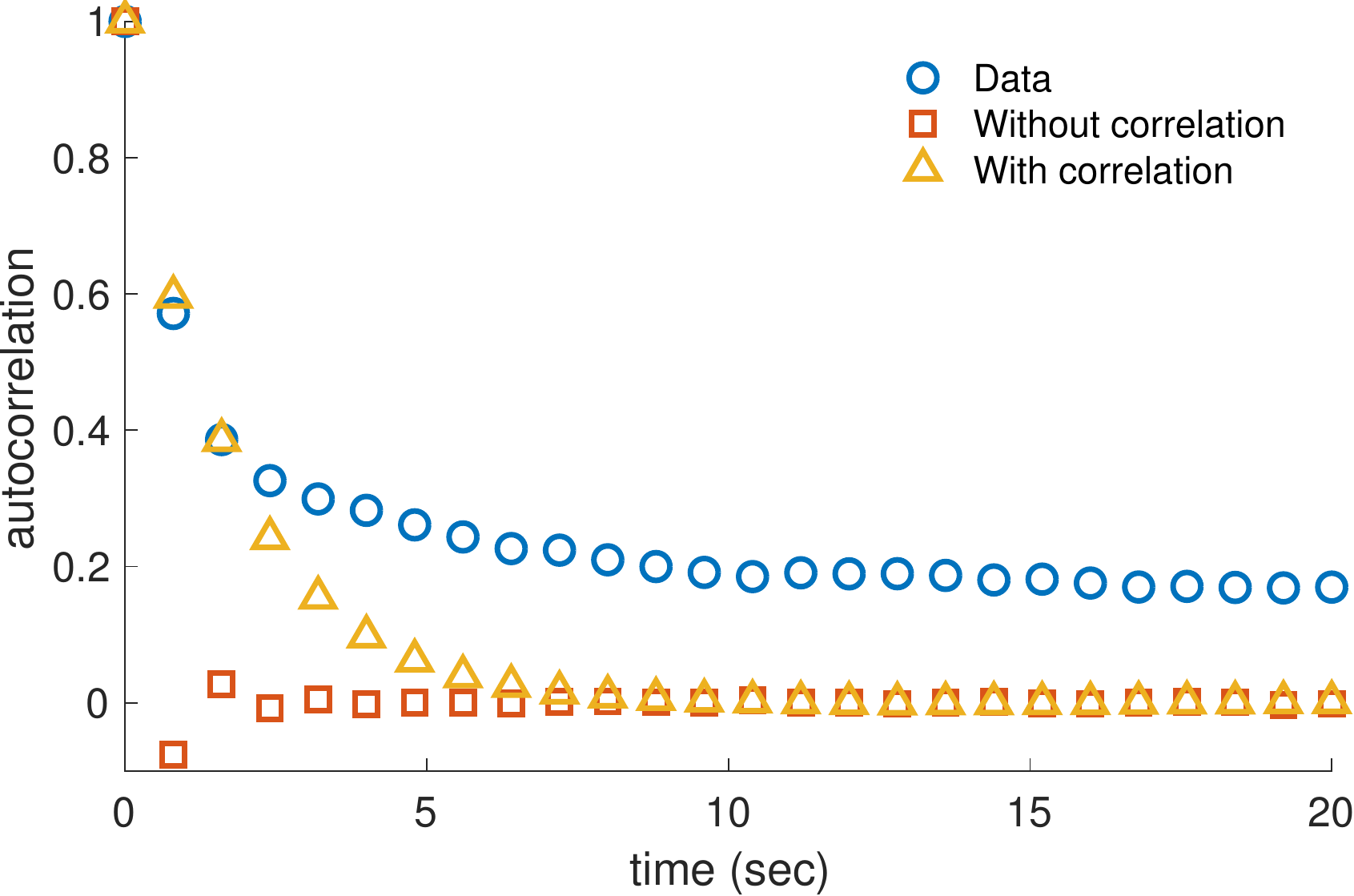}}
\caption{The autocorrelation as function of time difference in the signal, for the data (blue circles), the zeroth-order model  (red squares), and the first-order model (yellow triangles). The first-order model correctly approximates the auto-correlation up to $\approx$ 2 time steps (1.6 seconds). As expected, the zeroth-order model yields a flat autocorrelation of $\approx 0$.}
\end{figure*}

\clearpage
\section*{Uncorrelated null model}
Shuffling the speeds without considering the correlations shown in Fig. 3A produces trajectories deviating from the empirical distribution and scaling laws observed.

\begin{figure*}[h!]
\centering{\includegraphics[width=.48\textwidth]{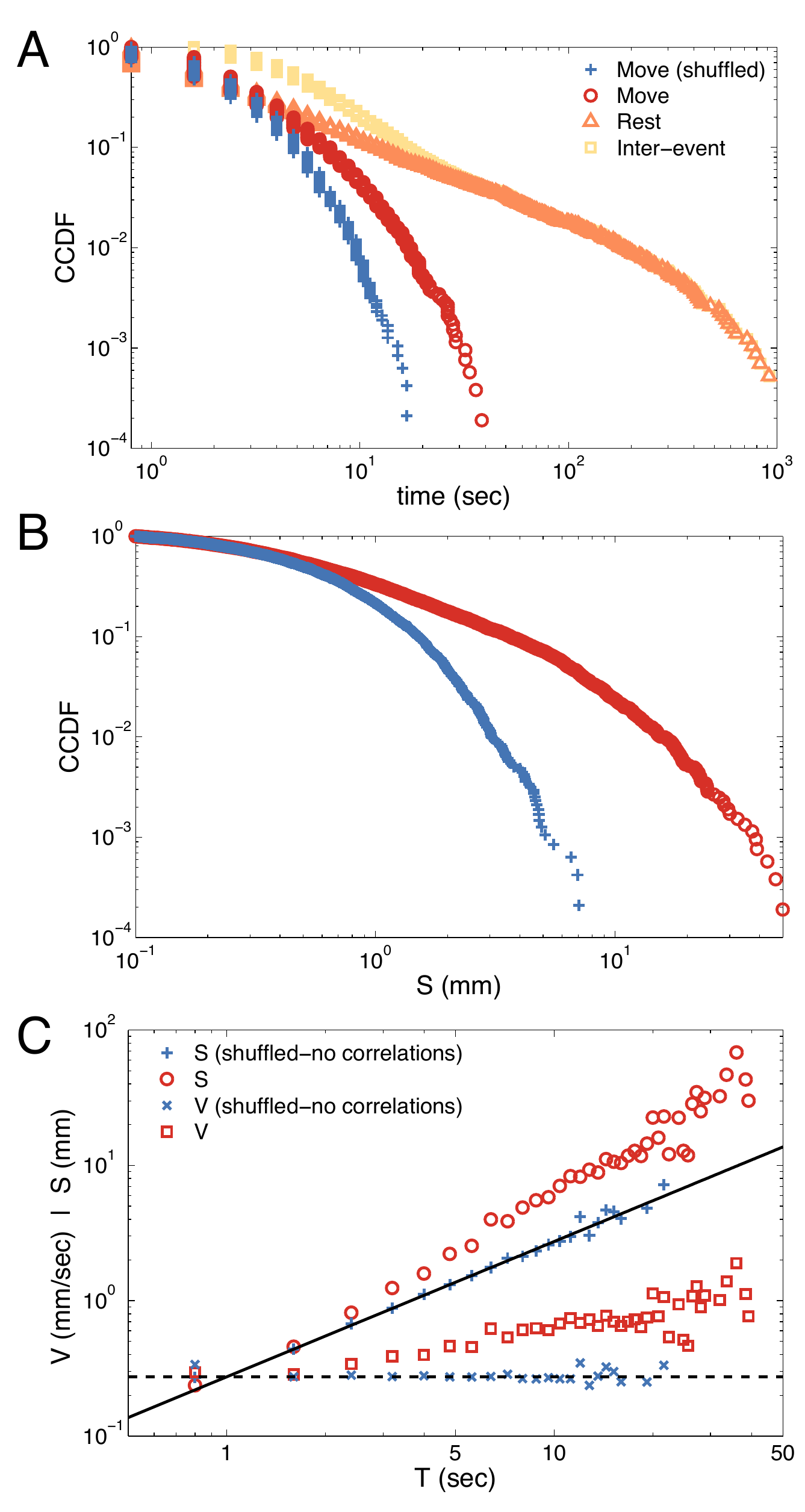}}
\caption{(A) The complementary cumulative distribution function of the events duration $T$, the inter-event time and rest time, for uncorrelated shuffled trajectories and real ants.
(B) The complementary cumulative distribution function of the events size $S$ for real ants and the uncorrelated null model. (C) The scaling of the events size and speed as a function of  events duration for the raw data as well as  for the uncorrelated null model shows no relevant differences. Continuous and dashed lines correspond to slopes of 1 and 0 (constant) respectively. Data from colony $C_{35\times 28}^2$.}
\end{figure*}
 
 \clearpage
\section*{Experimental results for all colonies}

Some of the experimental results of our paper are shown in the main text only for a single experimental condition. In this section we show the exact same analysis to a all colonies and sizes of the nets.

\begin{figure*}[h!]
\begin{tabular}{cc}
\raisebox{3.5cm}{$C_{35\times 28}^1$} \includegraphics[angle=0, width=0.35\textwidth]{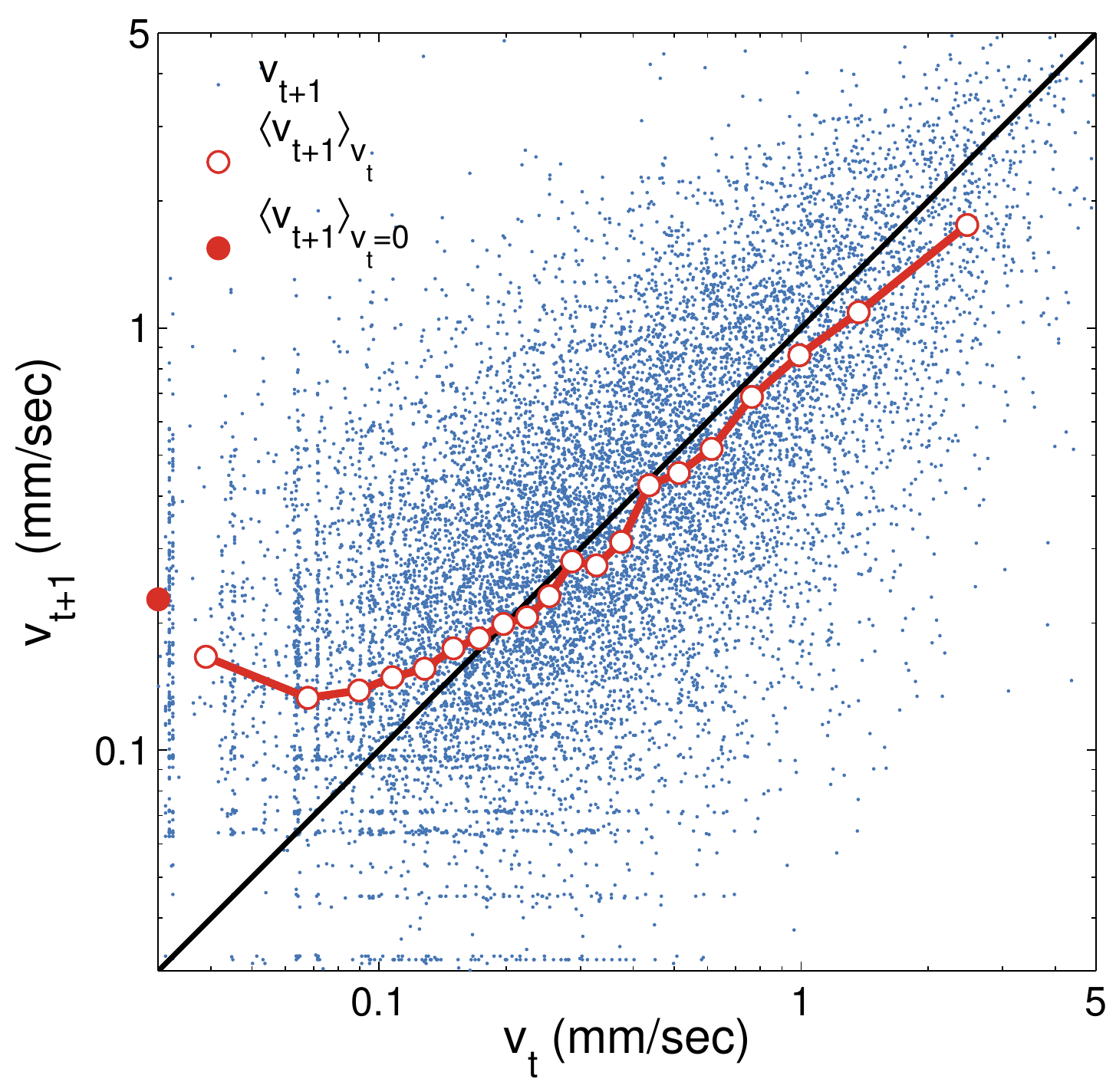}&
\raisebox{3.5cm}{$C_{55\times 44}^1$} \includegraphics[angle=0, width=0.35\textwidth]{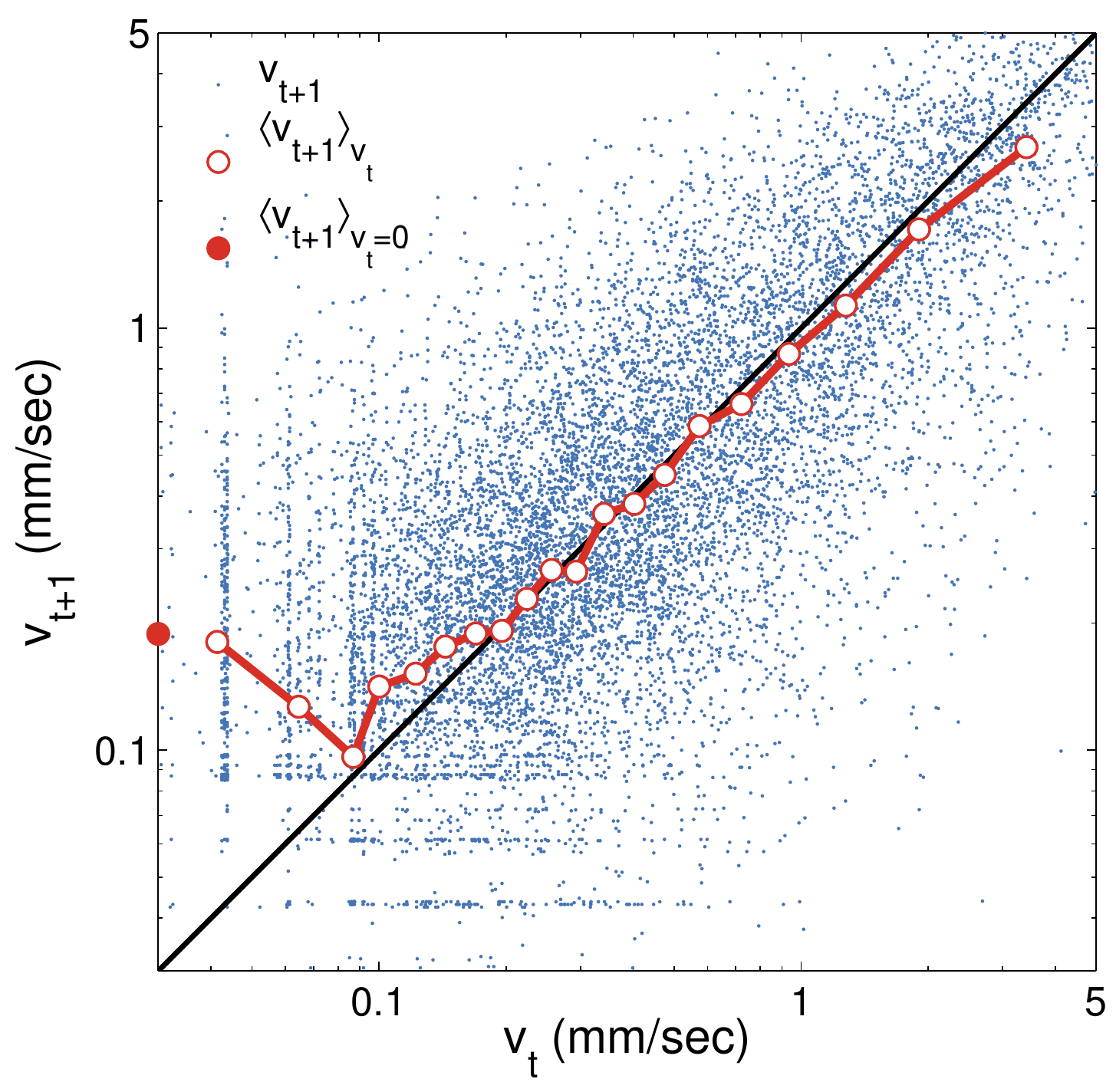}\\
\raisebox{3.5cm}{$C_{35\times 28}^2$} \includegraphics[angle=0, width=0.35\textwidth]{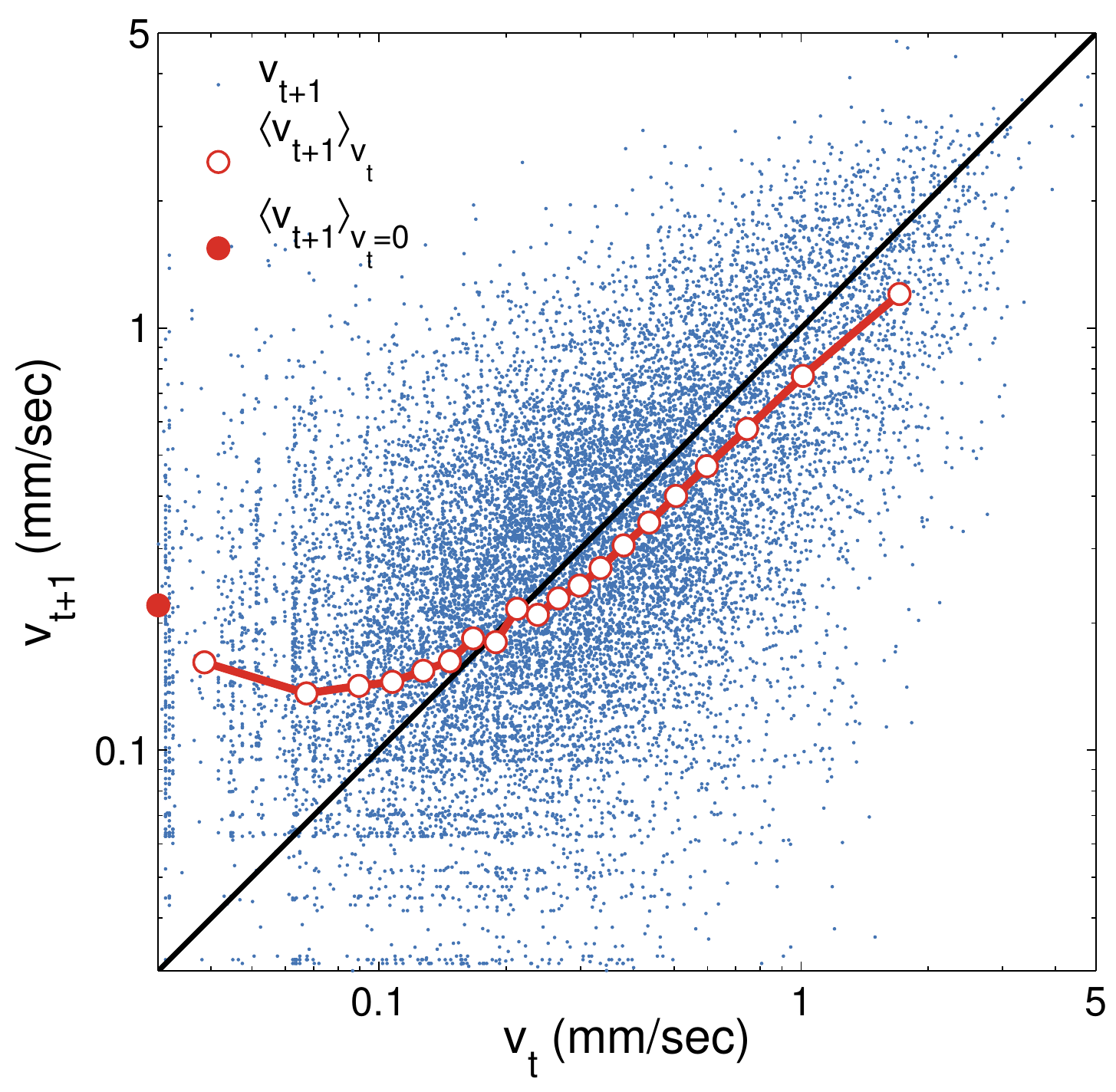}&
\raisebox{3.5cm}{$C_{55\times 44}^2$} \includegraphics[angle=0, width=0.35\textwidth]{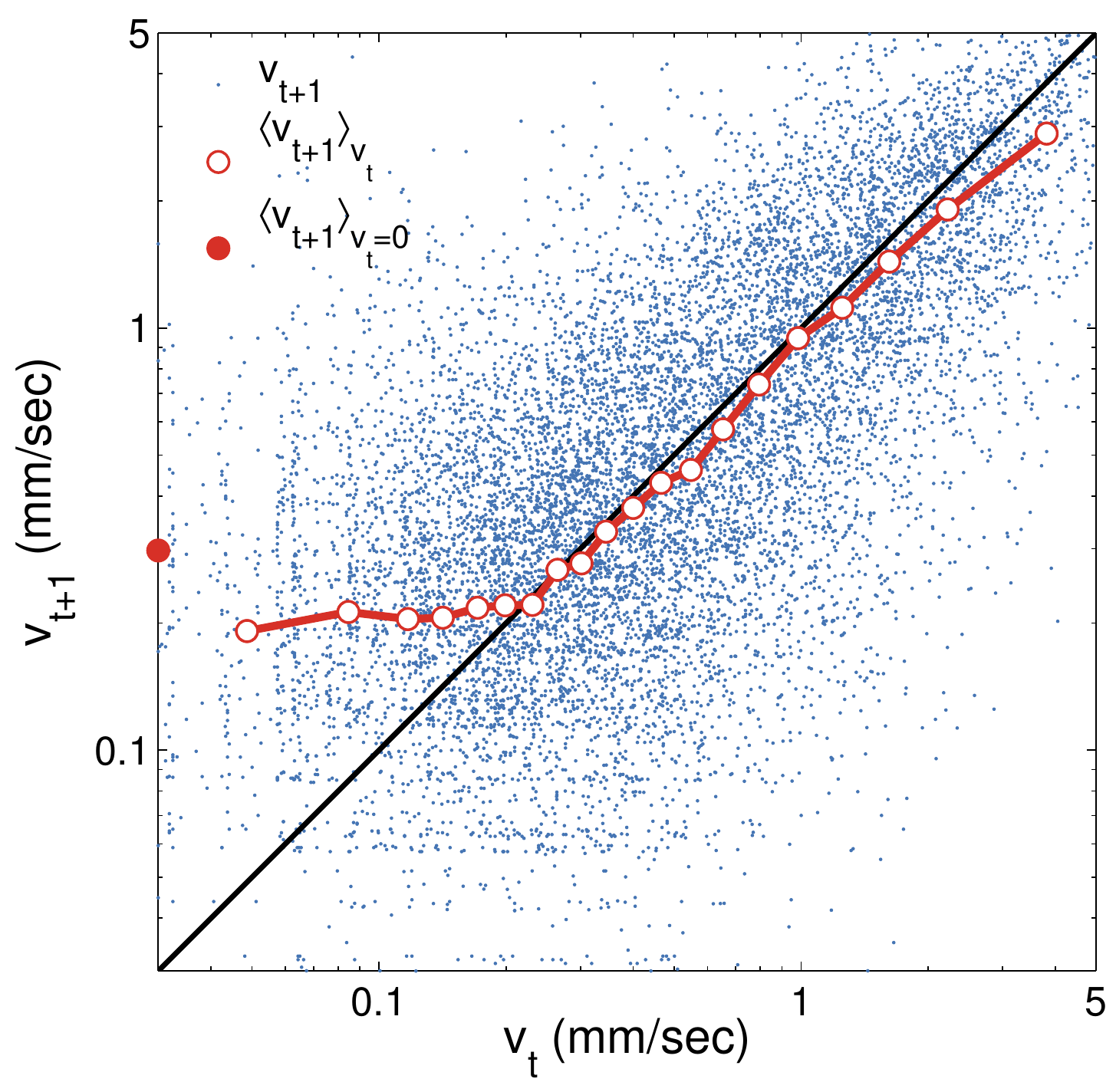}\\
\raisebox{3.5cm}{$C_{35\times 28}^3$} \includegraphics[angle=0, width=0.35\textwidth]{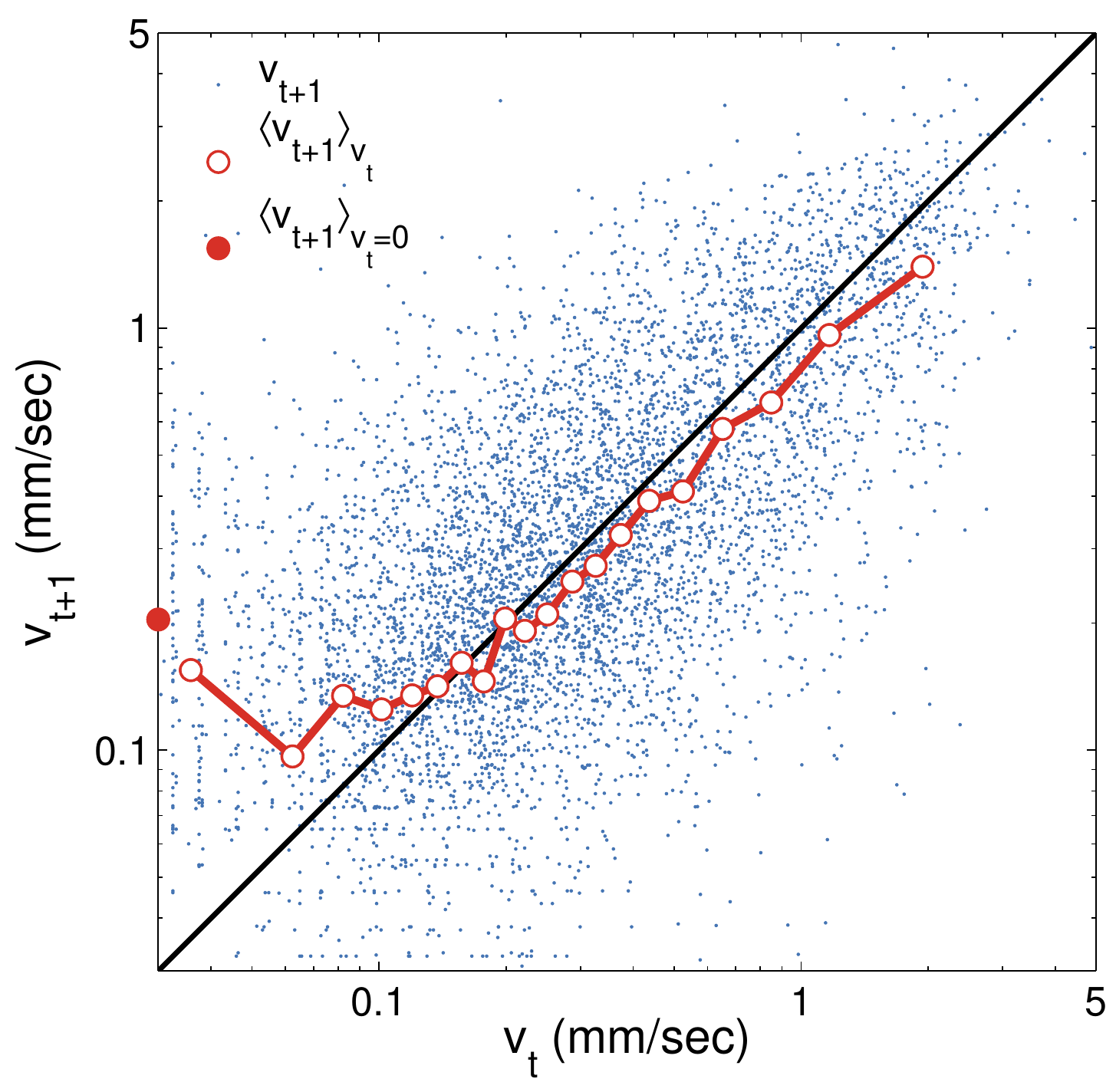}&
\raisebox{3.5cm}{$C_{55\times 44}^3$} \includegraphics[angle=0, width=0.35\textwidth]{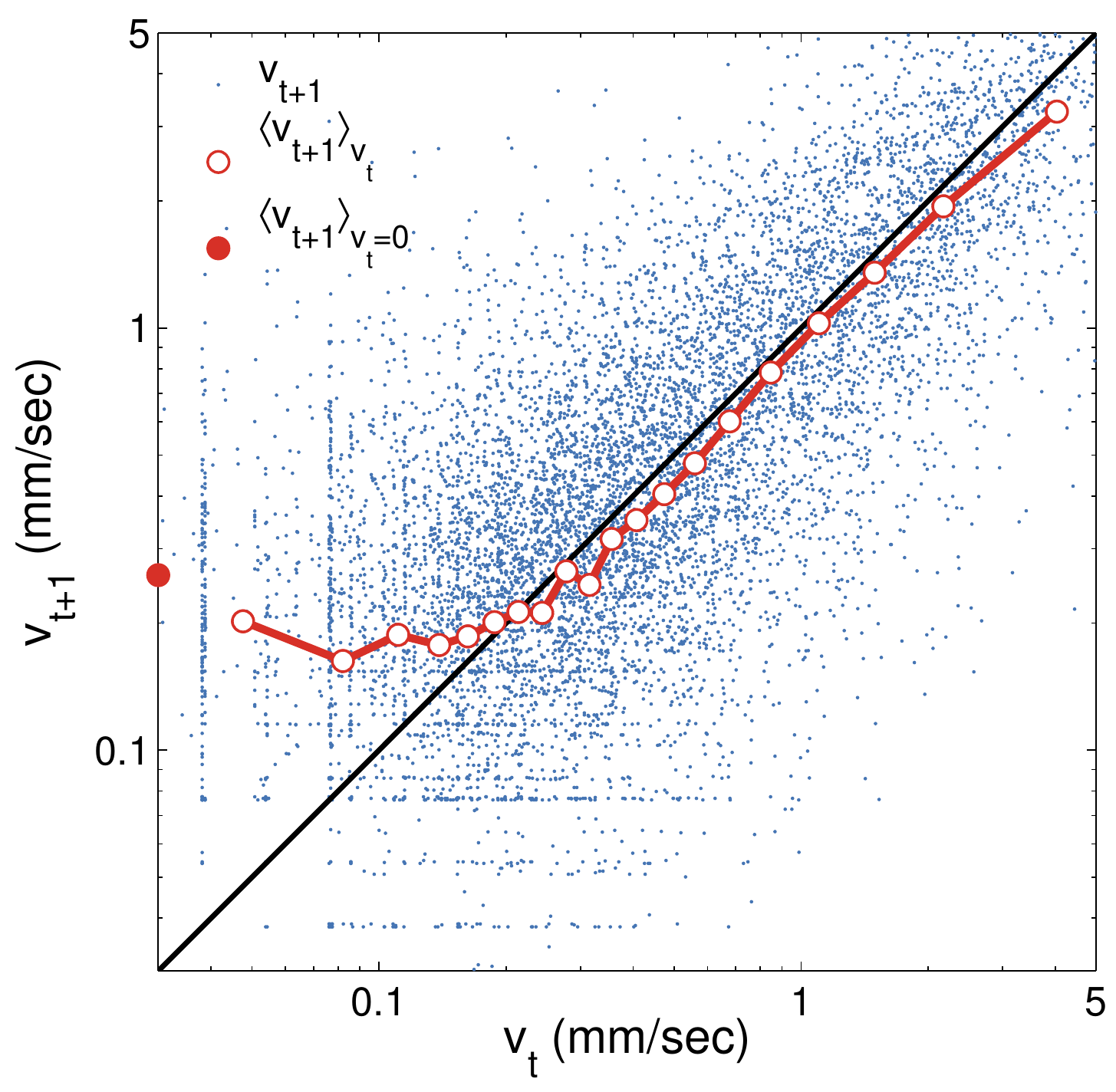}\\
\end{tabular}
\caption{{\bf Replicas of Fig.~3A.} Return map extracted from the raw data by plotting consecutive $v(t)$ samples (dots) and its binned average $f (v_{(t)})$  (circles and continuos line) over-imposed (please note the log axis). }
\label{}
\end{figure*}

\begin{figure*}[h!]
\begin{tabular}{cc}
\raisebox{2.5cm}{$C_{35\times 28}^1$} \includegraphics[angle=0, width=0.4\textwidth]{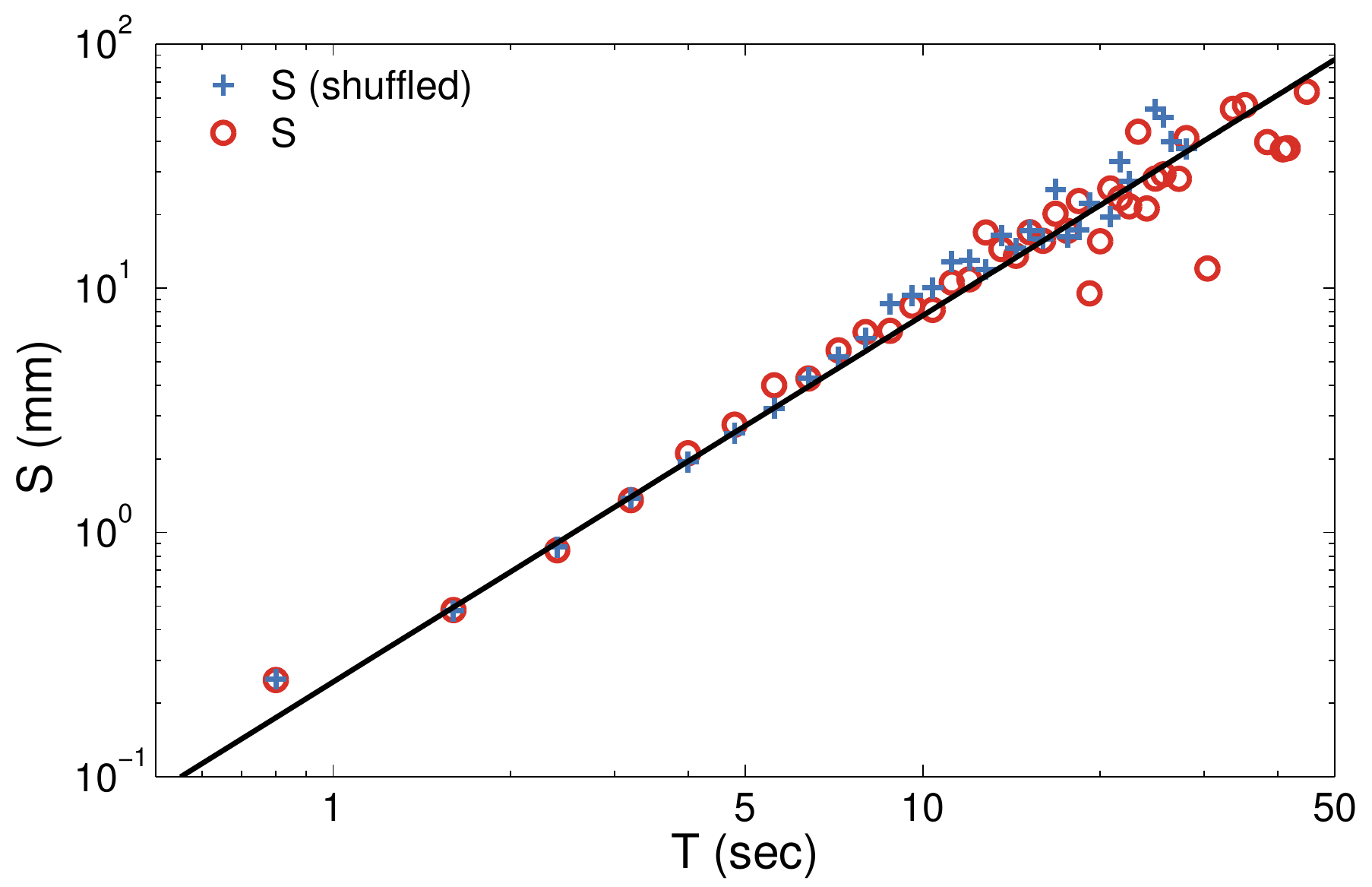}&
\raisebox{2.5cm}{$C_{55\times 44}^1$} \includegraphics[angle=0, width=0.4\textwidth]{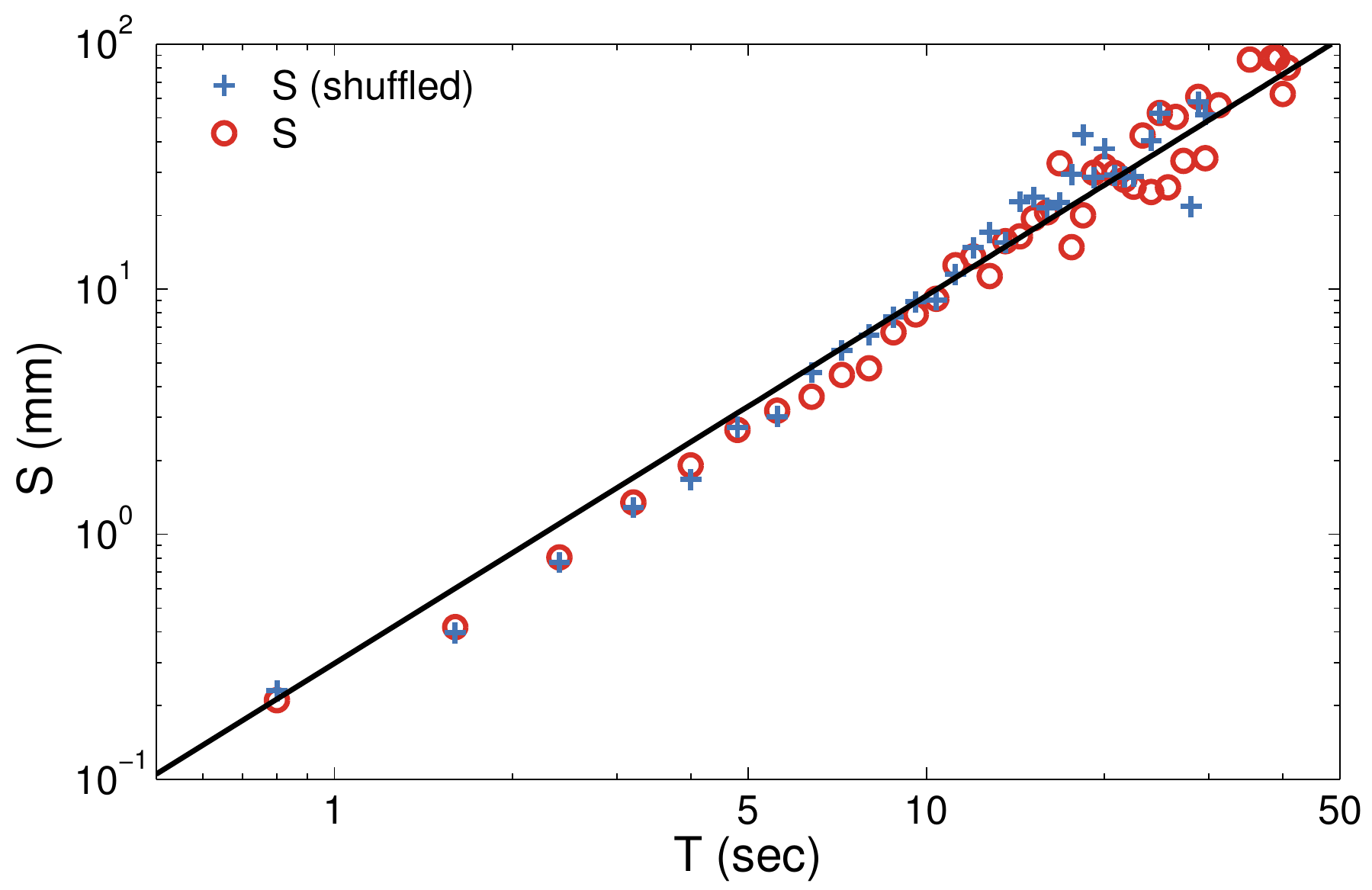}\\
\raisebox{2.5cm}{$C_{35\times 28}^2$} \includegraphics[angle=0, width=0.4\textwidth]{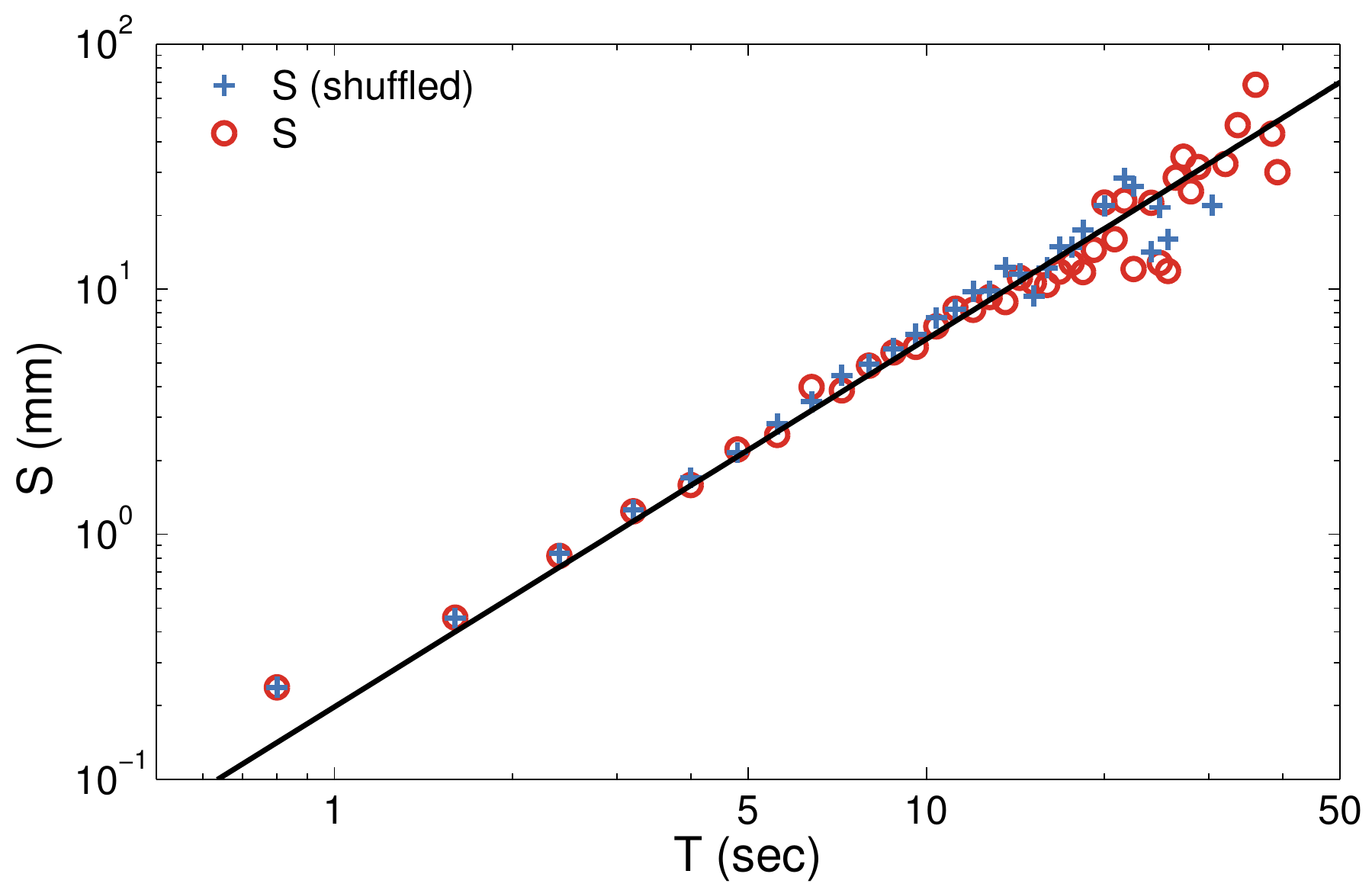}&
\raisebox{2.5cm}{$C_{55\times 44}^2$} \includegraphics[angle=0, width=0.4\textwidth]{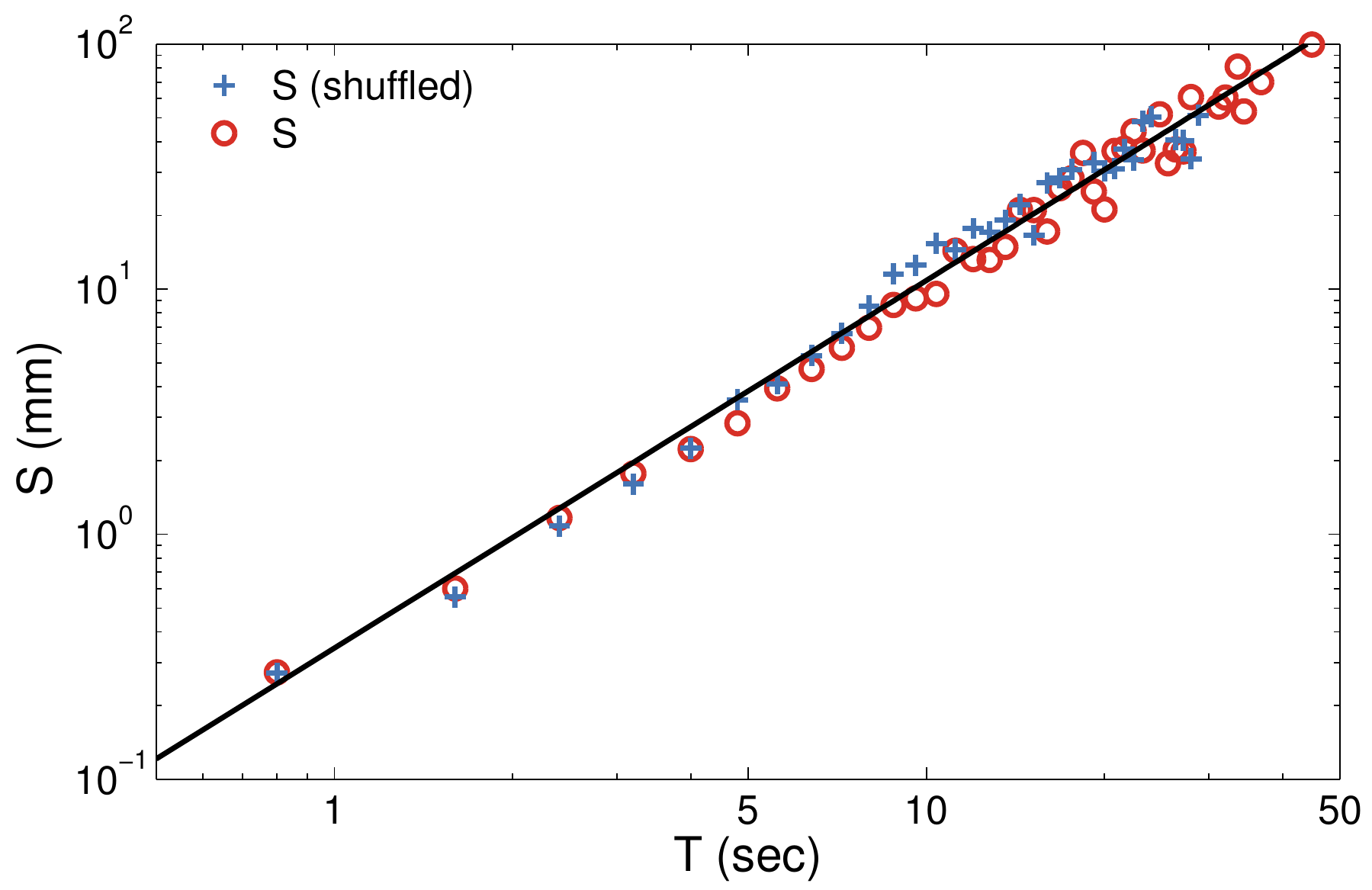}\\
\raisebox{2.5cm}{$C_{35\times 28}^3$} \includegraphics[angle=0, width=0.4\textwidth]{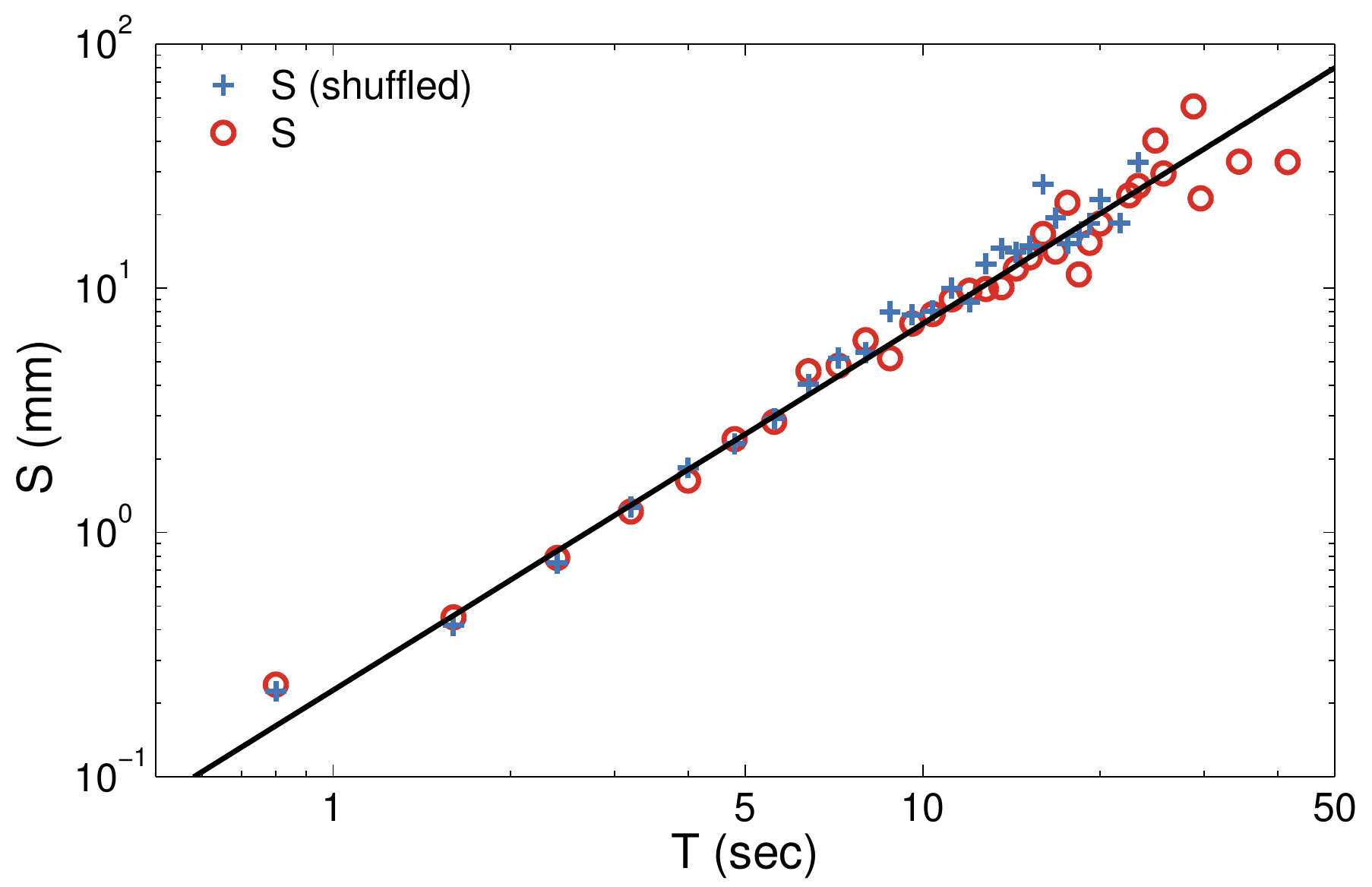}&
\raisebox{2.5cm}{$C_{55\times 44}^3$} \includegraphics[angle=0, width=0.4\textwidth]{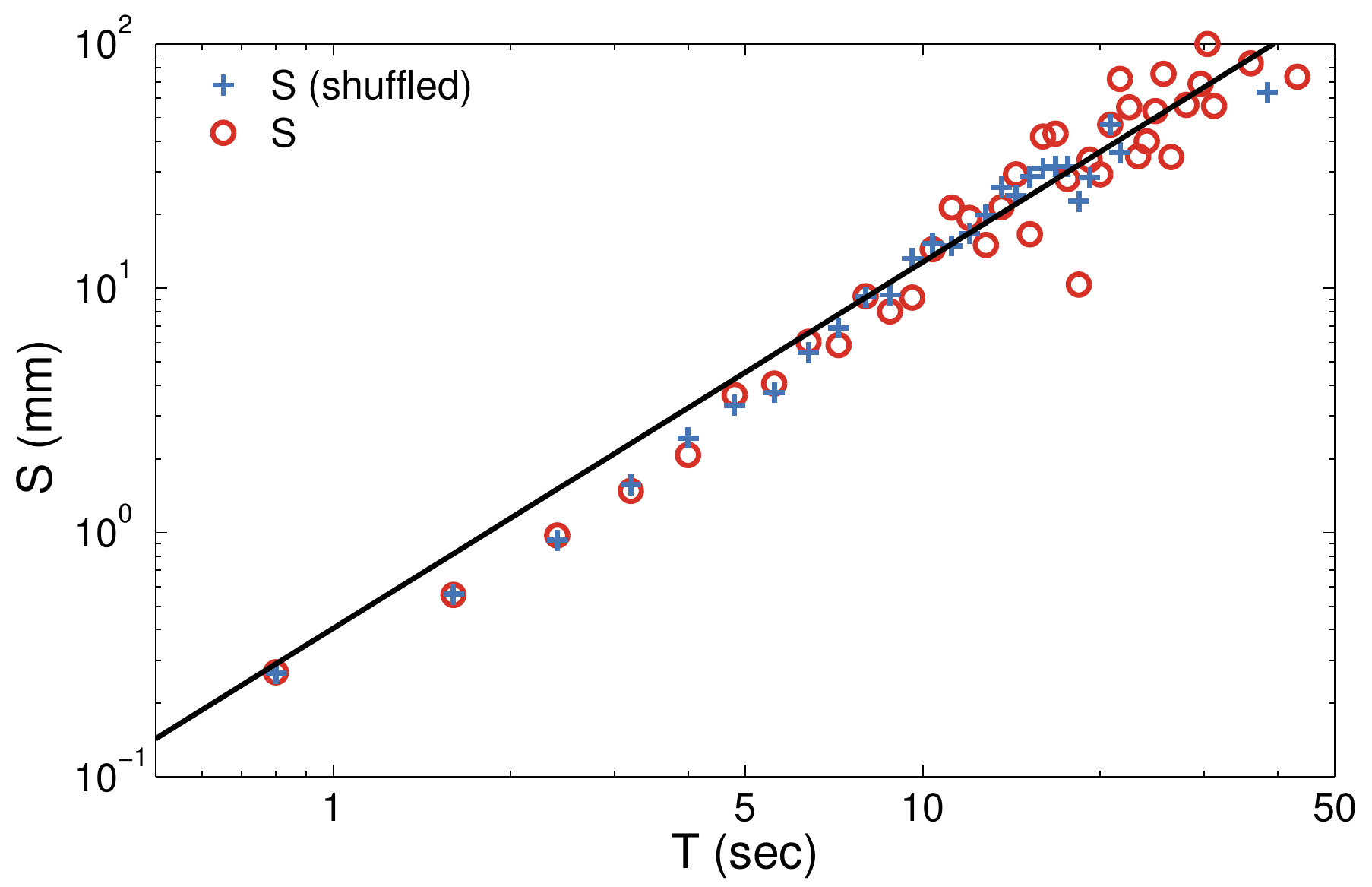}\\
\end{tabular}
\caption{{\bf Replicas of Fig.~4C.} Real and null model trajectories exhibit very similar scaling statistics. The scaling of the events size and speed as a function of  events duration for the raw data as well as for the null model shows no relevant differences. The solid line correspond to a slope of 1/2. }
\label{}
\end{figure*}

\begin{figure*}[h!]
\begin{tabular}{cc}
\raisebox{2.5cm}{$C_{35\times 28}^1$} \includegraphics[angle=0, width=0.4\textwidth]{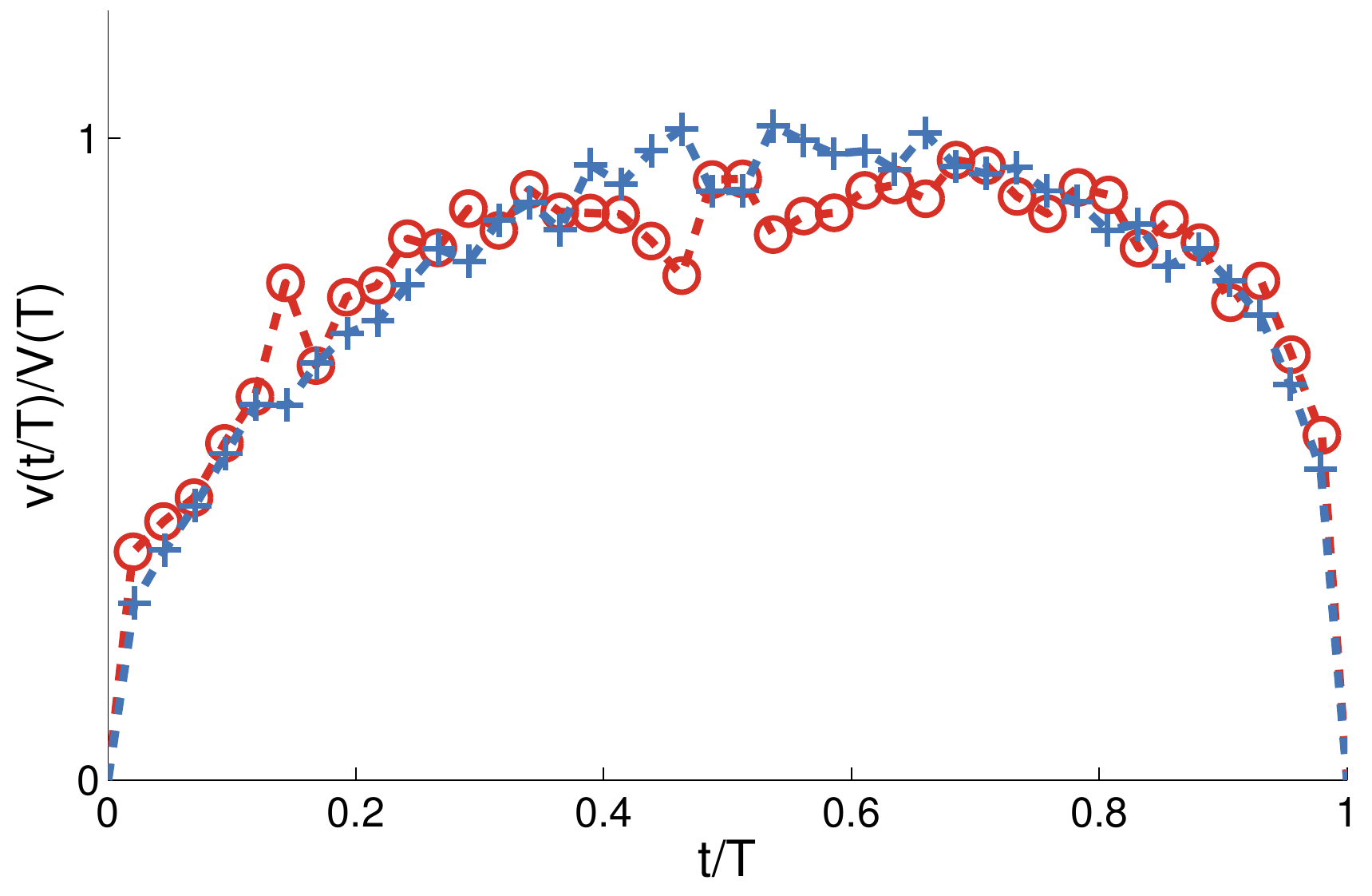}&
\raisebox{2.5cm}{$C_{55\times 44}^1$} \includegraphics[angle=0, width=0.4\textwidth]{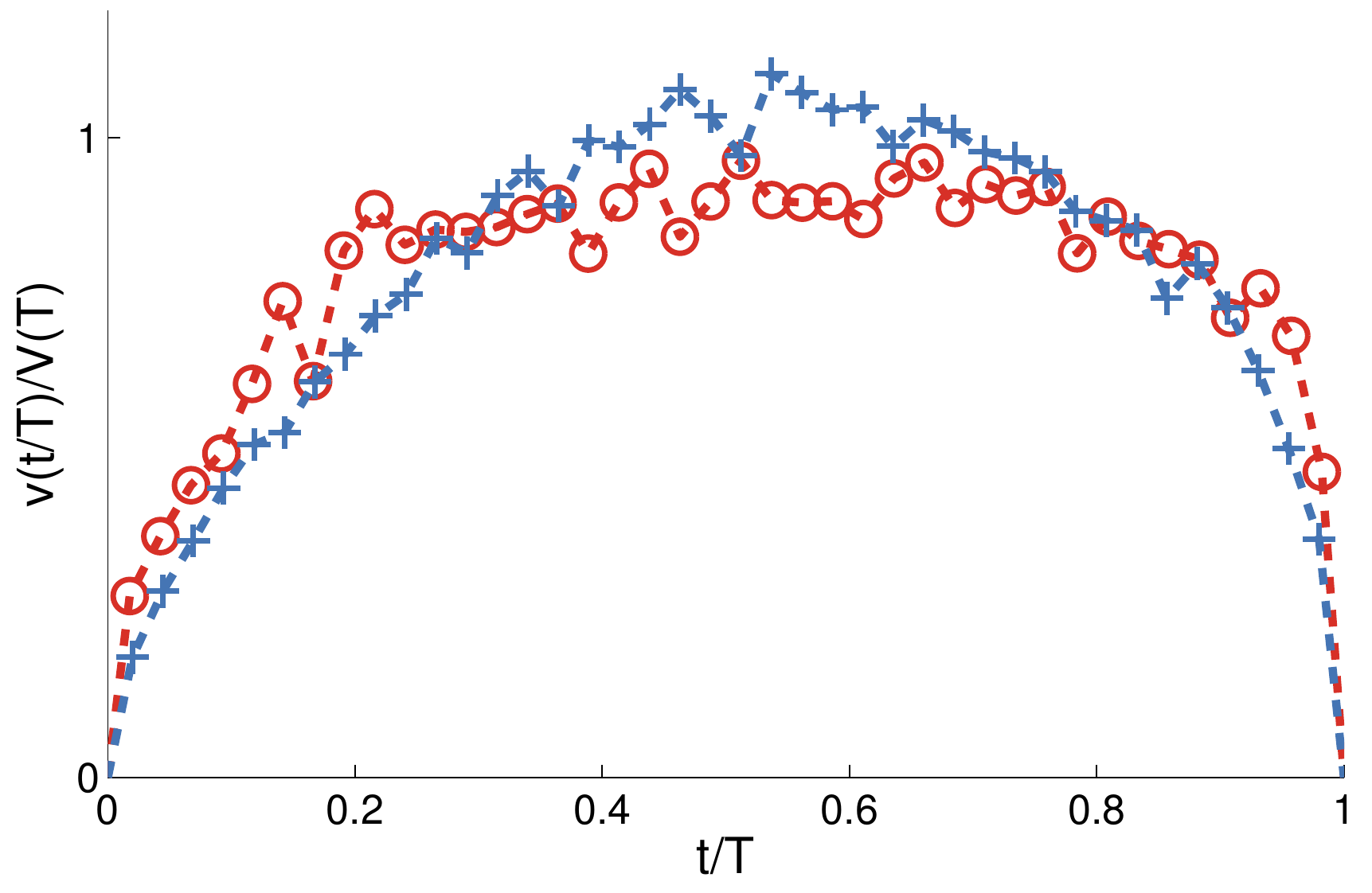}\\
\raisebox{2.5cm}{$C_{35\times 28}^2$} \includegraphics[angle=0, width=0.4\textwidth]{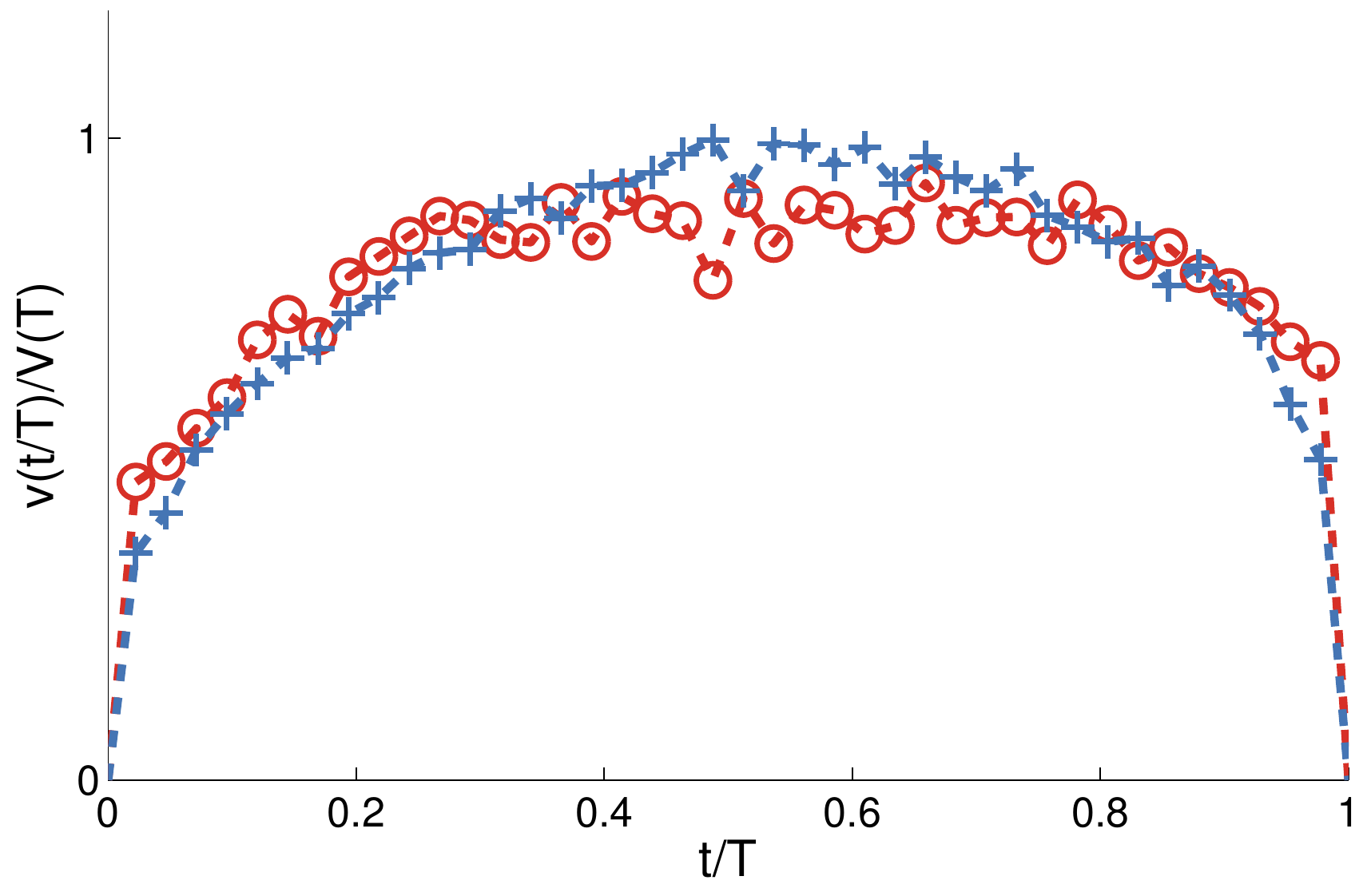}&
\raisebox{2.5cm}{$C_{55\times 44}^2$} \includegraphics[angle=0, width=0.4\textwidth]{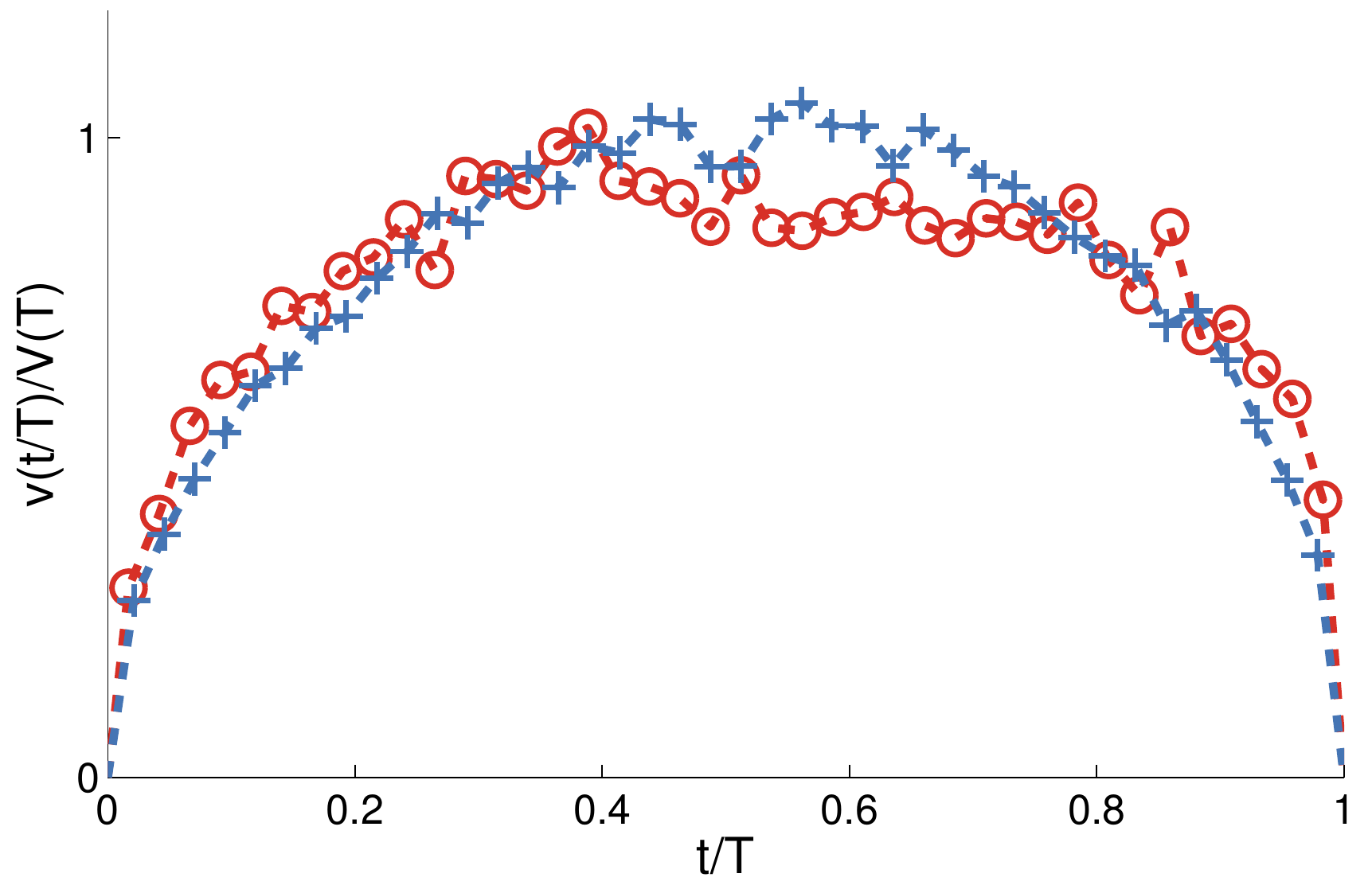}\\
\raisebox{2.5cm}{$C_{35\times 28}^3$} \includegraphics[angle=0, width=0.4\textwidth]{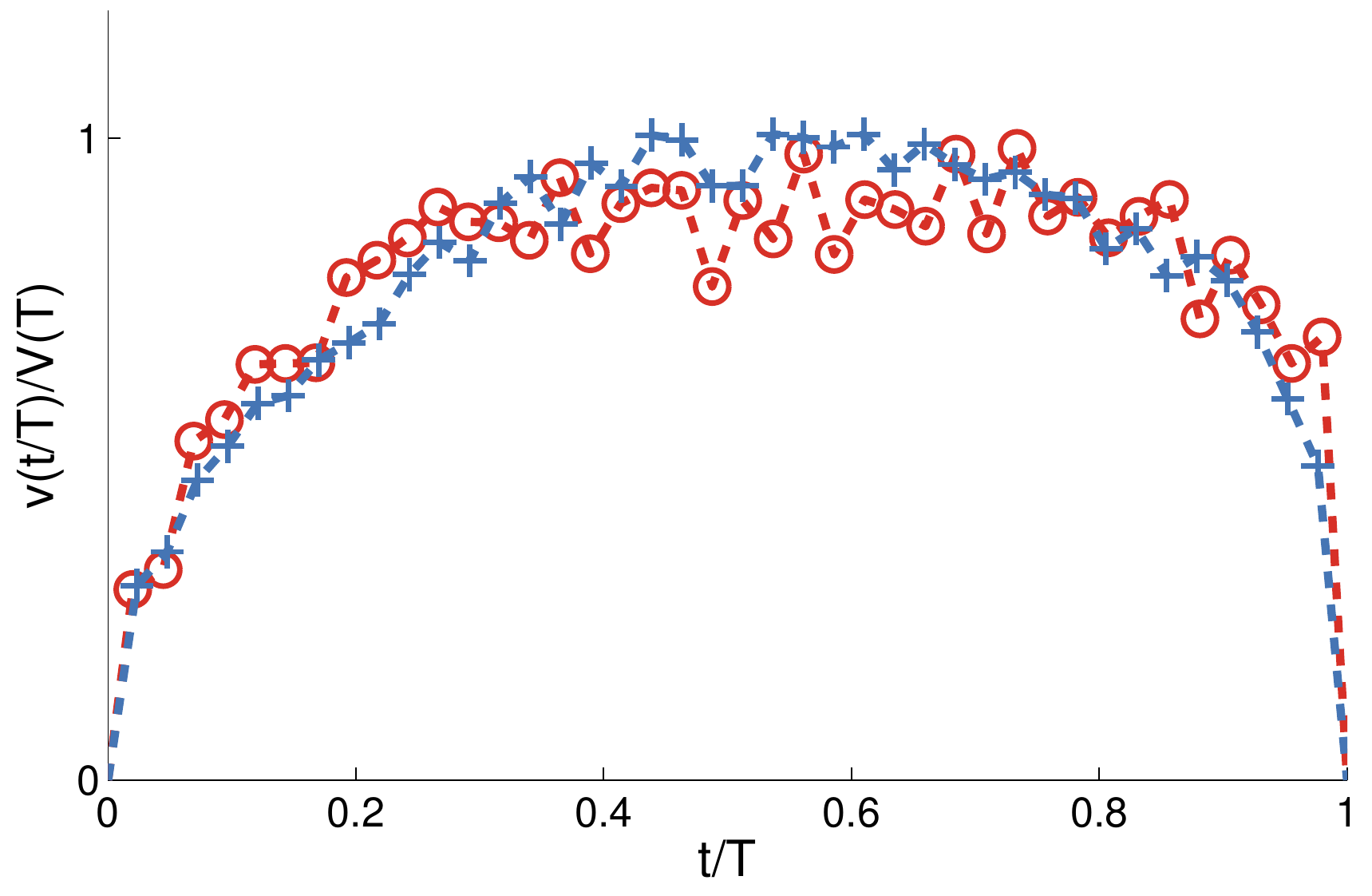}&
\raisebox{2.5cm}{$C_{55\times 44}^3$} \includegraphics[angle=0, width=0.4\textwidth]{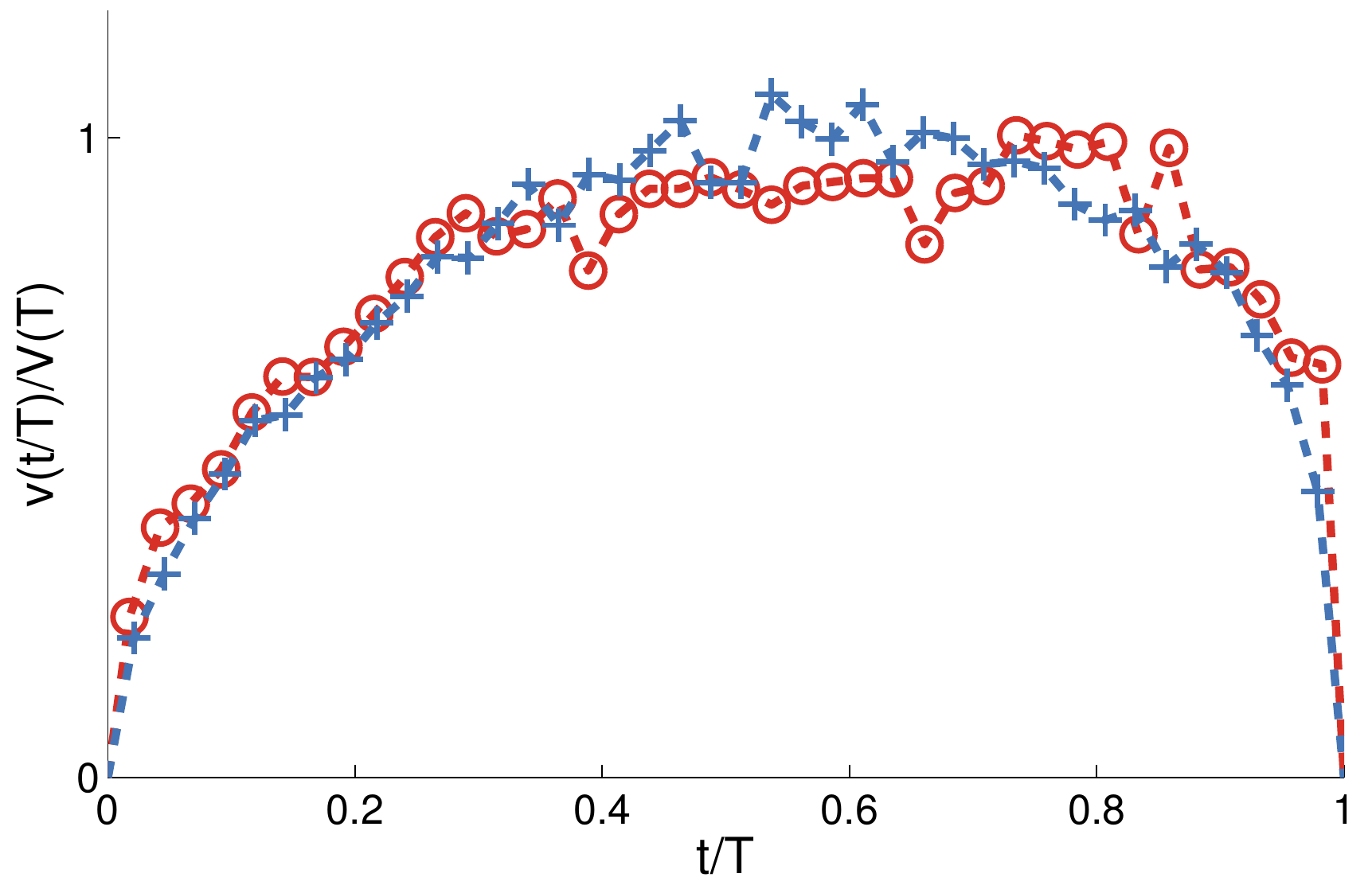}\\
\end{tabular}
\caption{{\bf Replicas of Fig.~5.} The normalized average speed profile for shuffled trajectories (blue plus signs) and real ants (red circles). 
After rescaling, all the events profiles collapse to an universal function that presents a plateau close to the average speed for intermediate time. The same collapse can be observed in the synthetic trajectories, however the shape of the curve is less flat. We notice that the shuffled trajectories also  reproduce the skew of the curve.}
\label{}
\end{figure*}

\begin{figure*}[h!]
\begin{tabular}{cc}
\raisebox{2.5cm}{$C_{35\times 28}^1$} \includegraphics[angle=0, width=0.4\textwidth]{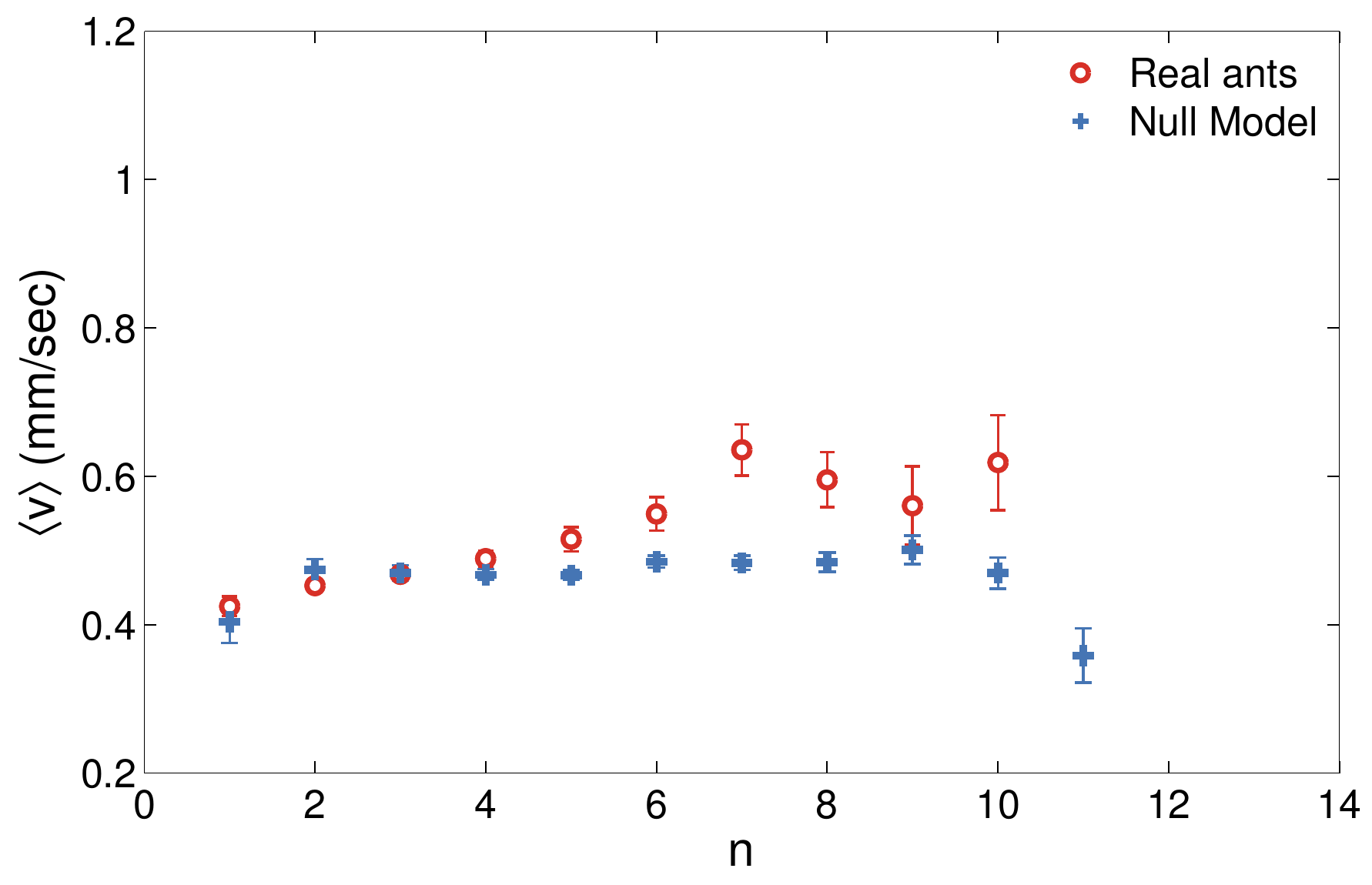}&
\raisebox{2.5cm}{$C_{55\times 44}^1$} \includegraphics[angle=0, width=0.4\textwidth]{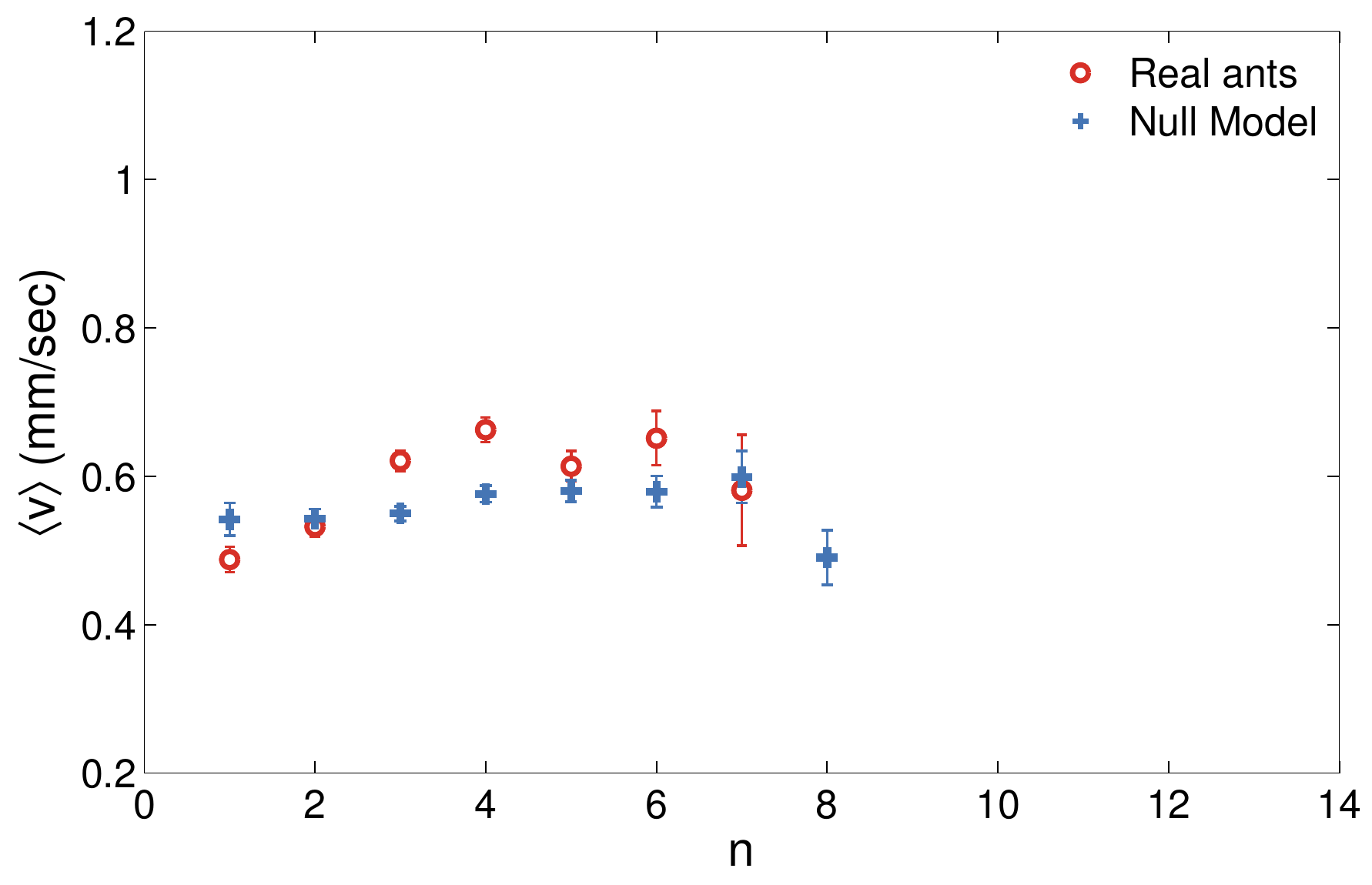}\\
\raisebox{2.5cm}{$C_{35\times 28}^2$} \includegraphics[angle=0, width=0.4\textwidth]{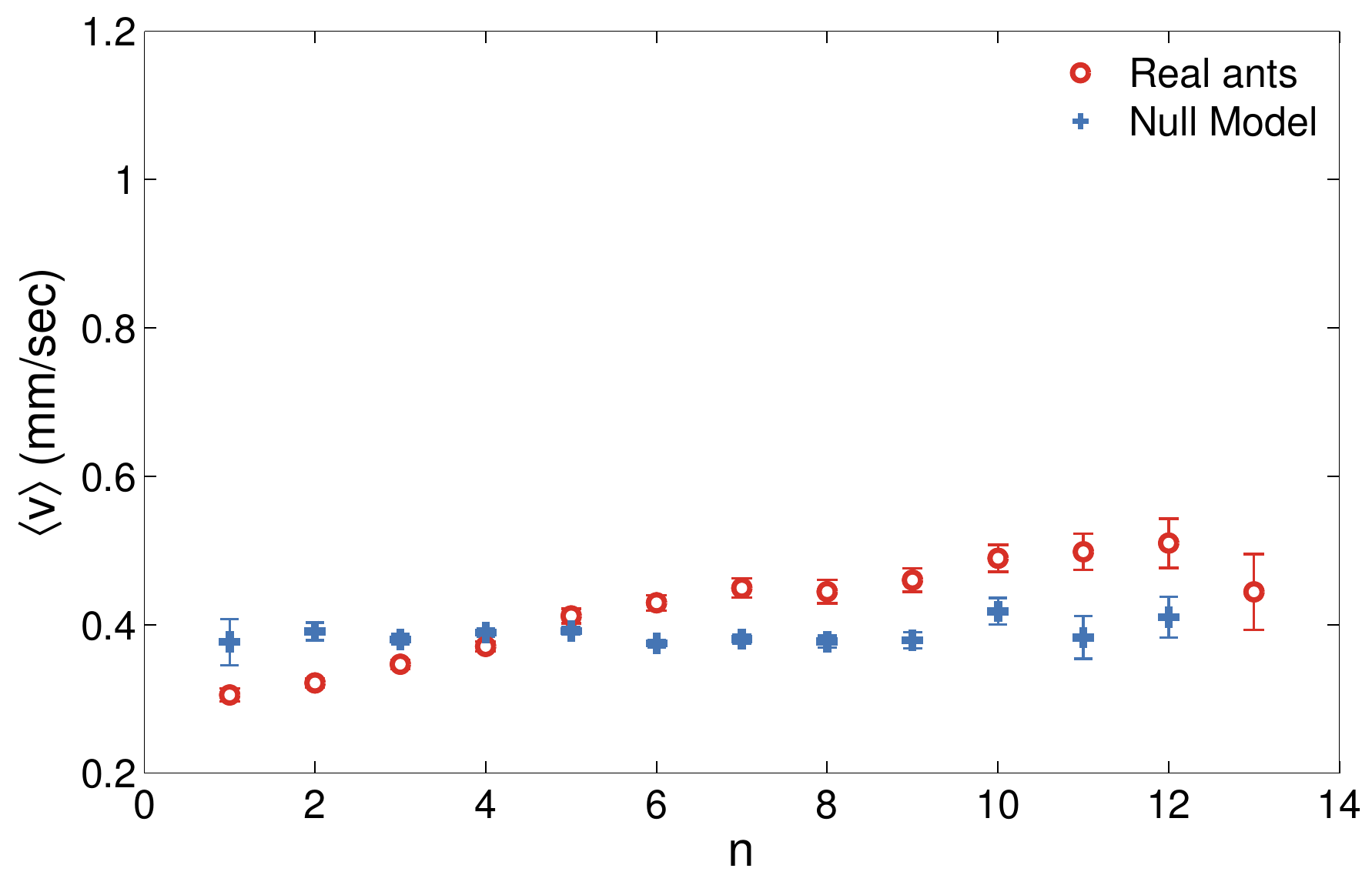}&
\raisebox{2.5cm}{$C_{55\times 44}^2$} \includegraphics[angle=0, width=0.4\textwidth]{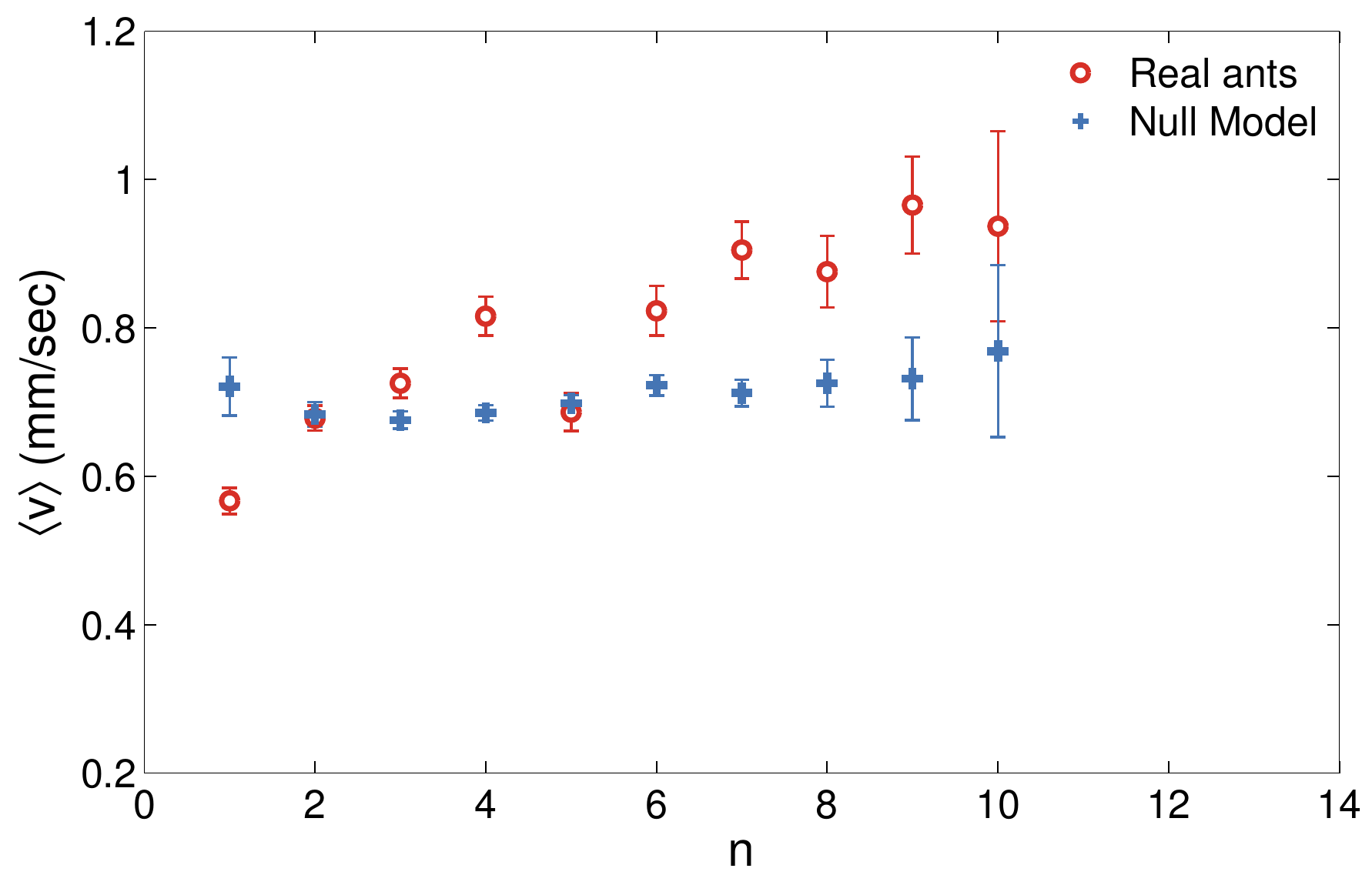}\\
\raisebox{2.5cm}{$C_{35\times 28}^3$} \includegraphics[angle=0, width=0.4\textwidth]{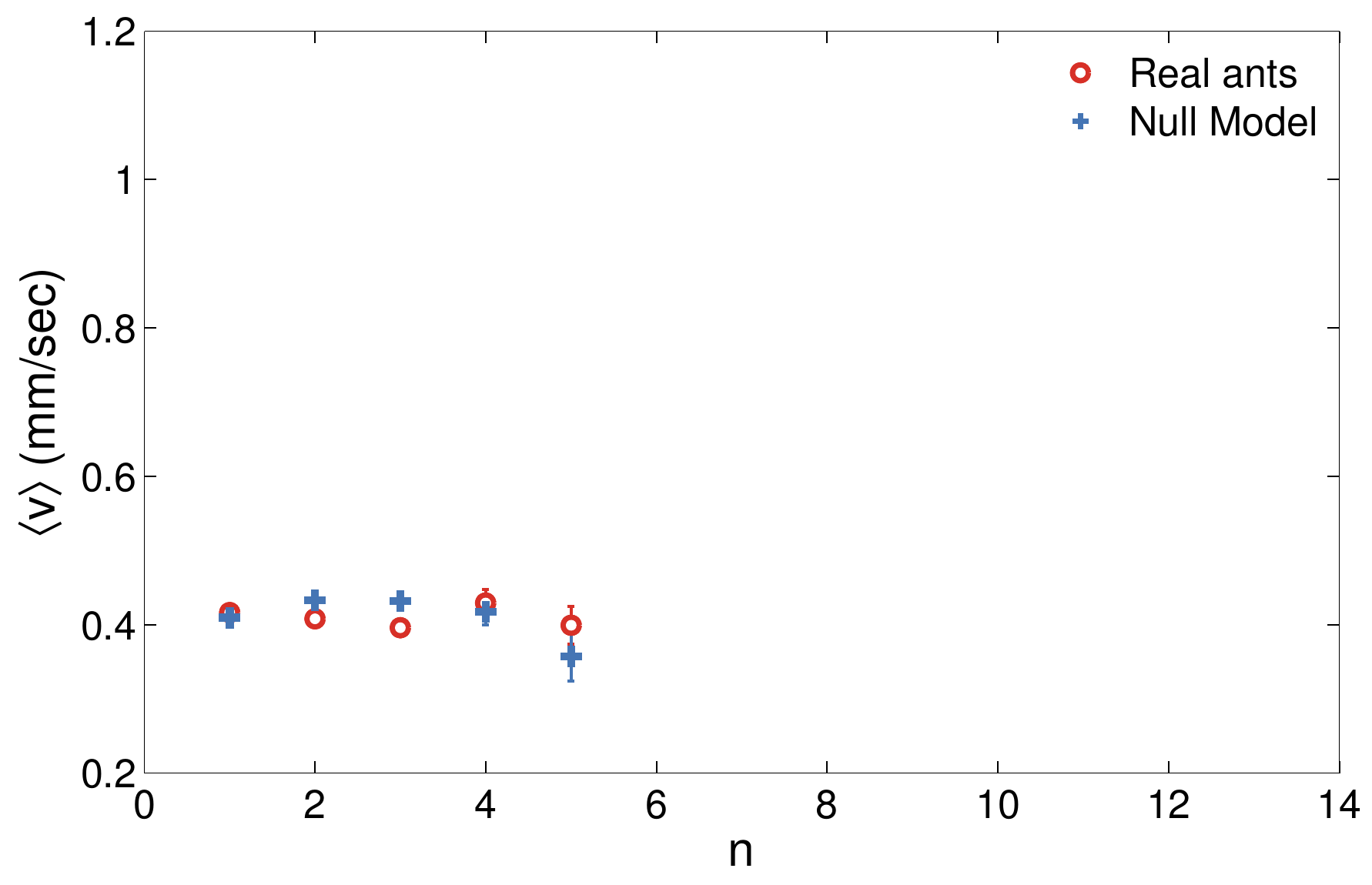}&
\raisebox{2.5cm}{$C_{55\times 44}^3$} \includegraphics[angle=0, width=0.4\textwidth]{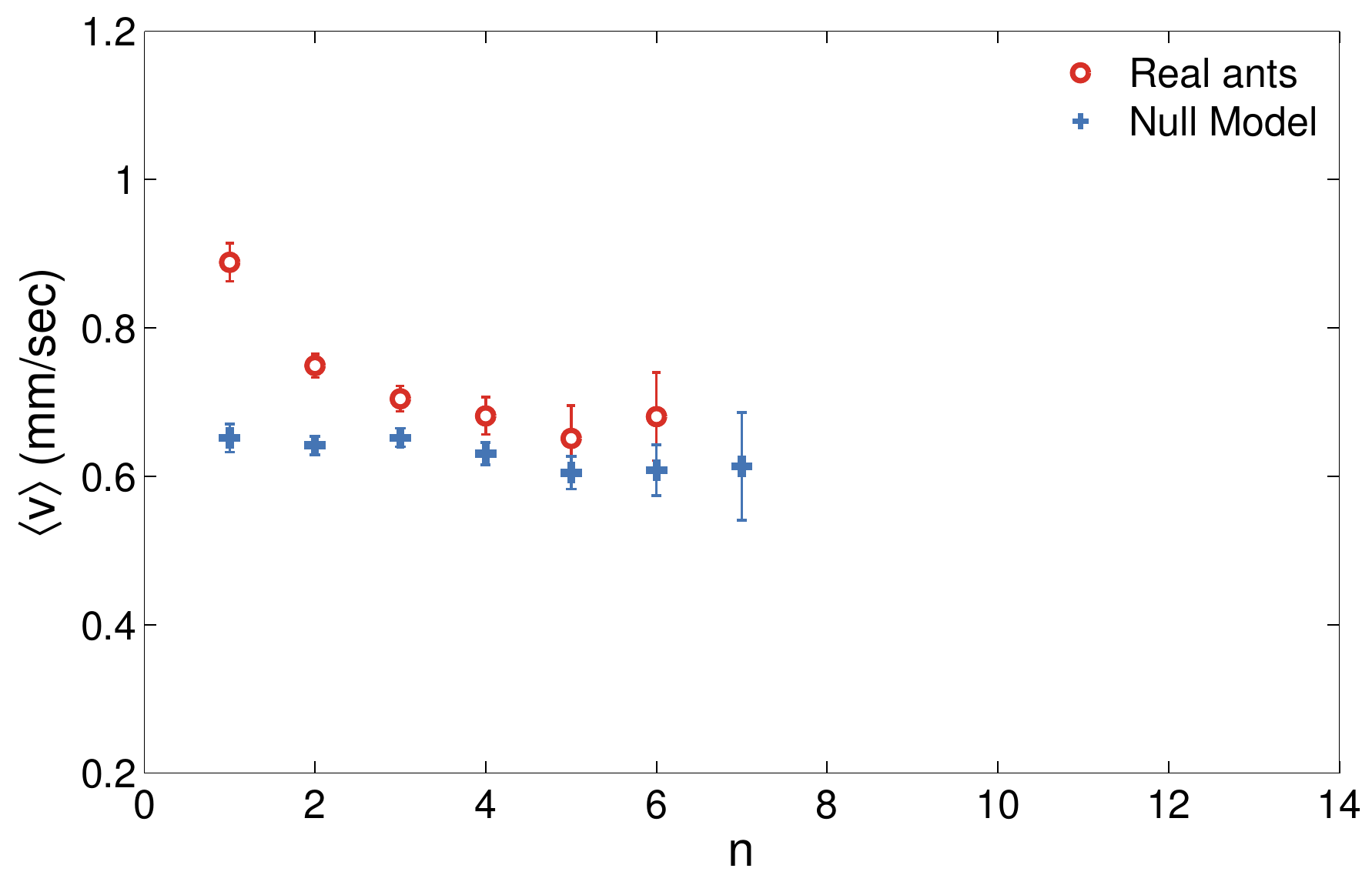}\\
\end{tabular}
\caption{{\bf Replicas of Fig.~6B.} The null model shows that growth of speed with $N$, observed for colonies $C^1$ and $C^2$, is not due to the statistics of single ant' behavior. }
\label{}
\end{figure*}

\clearpage
\section*{Dependence of individual and collective scaling on the colony and nest sizes.}
Our work used the data of three ant colonies with activities recorded in artificial nests of different sizes. The three colonies $C^1$, $C^2$, and $C^3$ were formed by 121, 92, 67 ants respectively at the time of the experiment. The experiment was run in nests of either $35\times28$mm or $55\times44$mm. Here we provide information on how the number of ants in each colony and the size affects the scaling effects observed.

\begin{figure*}[h!]
\begin{tabular}{cc}
\includegraphics[angle=0, width=0.4\textwidth]{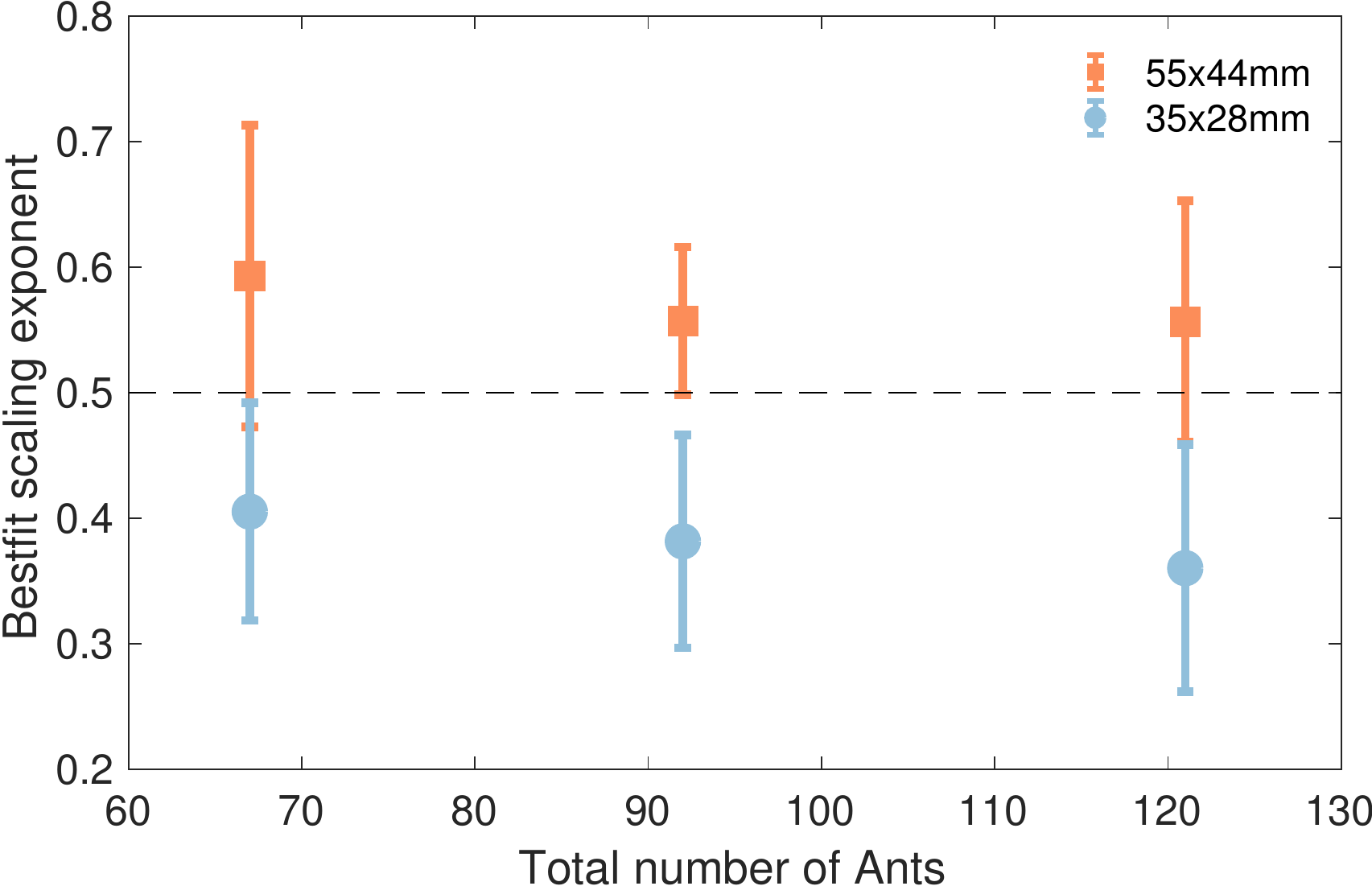}&
\includegraphics[angle=0, width=0.4\textwidth]{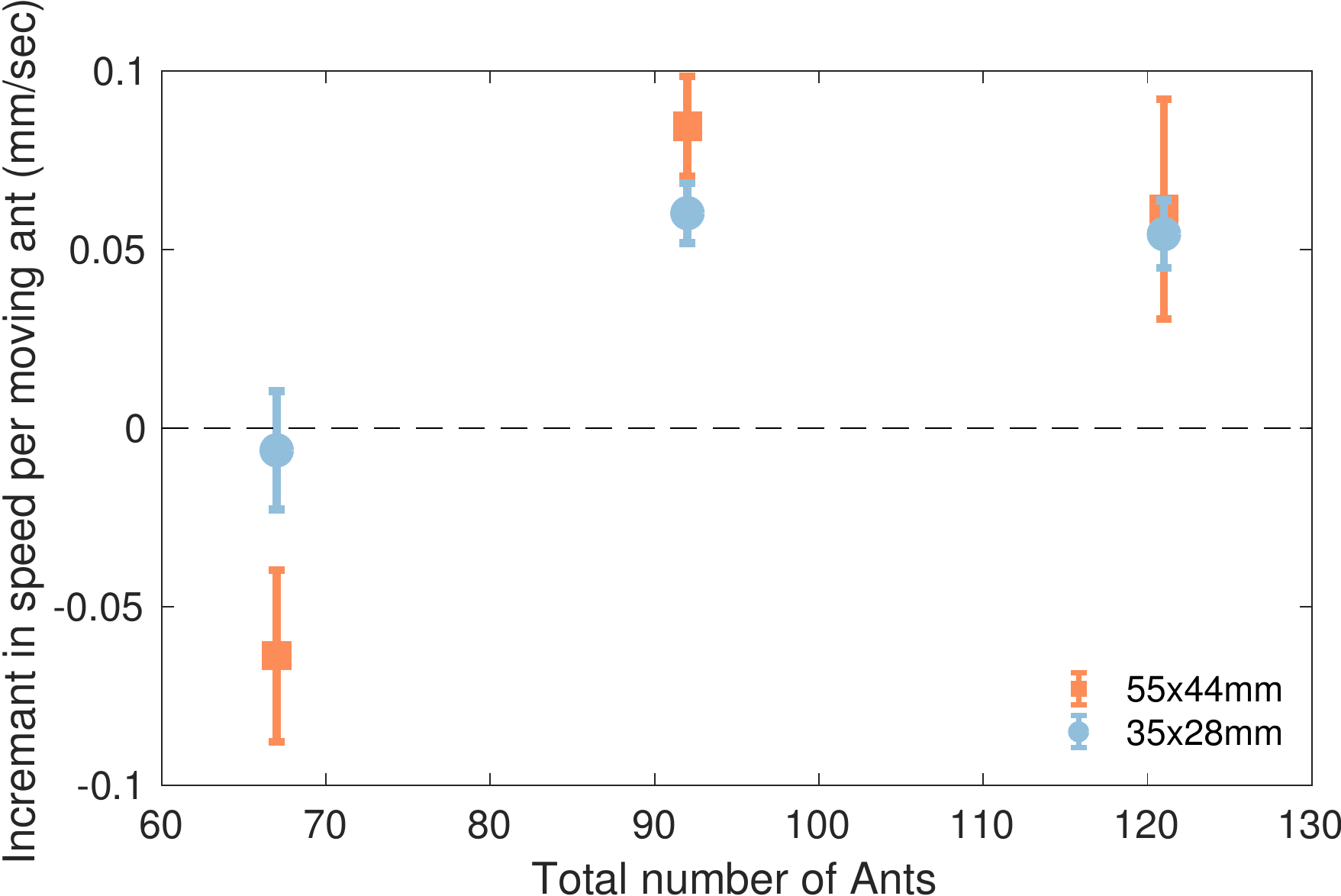}\\
\end{tabular}
\caption{
{\bf (Left)} The values of the slopes associated experimental best-fits of the curves proposed in Fig.~S4 for the different experimental setups does not depend on the number of ants, but varies with the nest's dimensions (the larger the nest, the larger the scaling exponent).
{\bf (Right)} The growing trend in the experimental data of Fig.~6A  is not observed for the colony with the smallest number of ants $C^3$, while is significantly positive for the two largest colonies $C^1$ and $C^2$ independently on the nest size.
}
\label{}
\end{figure*}

\end{document}